\documentclass[preprint]{aastex}

\newcommand{\degrees}{$^{\circ}$}
\newcommand{\Wms}{W~m$^{-2}$}
\newcommand{\Wkg}{W kg$^{-1}$}
\newcommand{\ms}{m~s$^{-1}$}
\newcommand{\kms}{km~s$^{-1}$}

\newcommand{\Prot}{\ensuremath{P_{\mathrm{rot}}}}
\newcommand{\kvis}{\ensuremath{\kappa_{\mathrm{vis}}}}

\newcommand{\Tint}{\ensuremath{T_{\mathrm{int}}}}
\newcommand{\Tirr}{\ensuremath{T_{\mathrm{irr}}}}
\newcommand{\tmag}{\ensuremath{\tau_{\mathrm{mag}}}}
\newcommand{\tmm}{\ensuremath{\tau_{\mathrm{mag,min}}}}

\begin{document}

\title{Three-Dimensional Atmospheric Circulation Models of HD 189733b \\
         and HD 209458b with Consistent Magnetic Drag and Ohmic Dissipation}

\author{Emily Rauscher\footnote{NASA Sagan Fellow}
  \\ \textit{Lunar and Planetary Laboratory, University of Arizona,
  \\ 1629 East University Blvd., Tucson, AZ 85721, USA}
  \\  and
  \\ Kristen Menou
  \\ \textit{Department of Astronomy, Columbia University,
  \\ 550 West 120th St., New York, NY 10027, USA}}
  
\begin{abstract}

We present the first three-dimensional circulation models for extrasolar gas giant atmospheres with geometrically and energetically consistent treatments of magnetic drag and ohmic dissipation.  Atmospheric resistivities are continuously updated and calculated directly from the flow structure, strongly coupling the magnetic effects with the circulation pattern.  We model the hot Jupiters HD 189733b ($T_{\mathrm{eq}}\approx1200$ K) and HD 209458b ($T_{\mathrm{eq}}\approx1500$ K) and test planetary magnetic field strengths from 0 to 30 G.  We find that even at $B=3$ G the atmospheric structure and circulation of HD 209458b are strongly influenced by magnetic effects, while the cooler HD 189733b remains largely unaffected, even in the case of $B=30$ G and super-solar metallicities.  Our models of HD~209458b indicate that magnetic effects can substantially slow down atmospheric winds, change circulation and temperature patterns, and alter observable properties.  These models establish that longitudinal and latitudinal hot spot offsets, day-night flux contrasts, and planetary radius inflation are interrelated diagnostics of the magnetic induction process occurring in the atmospheres of hot Jupiters and other similarly forced exoplanets.  Most of the ohmic heating occurs high in the atmosphere and on the day side of the planet, while the heating at depth is strongly dependent on the internal heat flux assumed for the planet, with more heating when the deep atmosphere is hot.  We compare the ohmic power at depth in our models, and estimates of the ohmic dissipation in the bulk interior (from general scaling laws), to evolutionary models that constrain the amount of heating necessary to explain the inflated radius of HD 209458b.  Our results suggest that deep ohmic heating can successfully inflate the radius of HD~209458b for planetary magnetic field strengths of $B\geq 3-10$ G.

\end{abstract}

\section{Introduction} \label{sec:intro}

Hot Jupiters are unlike any planet in our solar system and their atmospheric circulation exists in an entirely new regime \citep[for an extensive review, see][]{SCM}.  Orbiting within 0.1 AU of their host stars, these gas giants are subject to stellar irradiation levels $\sim 10^4$ times the flux Jupiter receives from our Sun.  This leads to a thick radiative zone atop their convective interiors and drives atmospheric winds that in many models reach or exceed the local sound speed \citep{Showman2009,DobbsDixon2012,Perna2012,RM12b}.  These exotic atmospheres have forced a reevaluation of which of the standard assumptions used in atmospheric dynamics may still be valid and whether new physical processes may need to be included in models in order to accurately determine their circulation patterns.  Although there are a growing number of observed planets, with measurements in expanding wavelength coverage, and an increasingly diverse set of methods by which to characterize their atmospheres, we will still never have as many photons from all exoplanets combined as we do from any single solar system planet \citep[see][for a review]{Seager2010}.  Nevertheless, these planets provide us with an exciting opportunity to extend the study of planetary atmospheres to strange new worlds.

It has recently been recognized that one of the novel processes that could affect a hot Jupiter's atmospheric circulation is due to the presence of the planet's own magnetic field.  The atmospheres of many of these planets are hot enough that they should be weakly thermally ionized (for pressures near the photosphere), with alkali metals providing the primary source of ions.\footnote{Note that at lower pressures (from nanobars to microbars) the atmosphere should be ionized by UV radiation from the star.  This is a distinct region from the pressure ranges covered by most general circulation models, with different dominant physical mechanisms and observable signatures.  See \citet{Koskinen2007} for an example of a circulation model in that regime.}  As the charged particles in the mostly neutral flow are advected around the planet by very fast winds, their circulation through the planetary magnetic field should result in the generation of a secondary component to the field and associated currents.  The effects on the atmosphere are predicted to be a bulk Lorentz force drag on the winds \citep{Perna2010a} and localized heating from ohmic dissipation of the currents \citep{Batygin2010,Perna2010b}.  In this paper we present the first circulation models to include these coupled effects.

There are several ways that magnetic effects may influence observable properties of hot Jupiters.  One of the most widely recognized is that the ohmic heating may provide the extra source of heating required to explain the unexpected large radii of some hot Jupiters \citep{Batygin2011,Laughlin2011,Huang2012,Wu2012}.  For a given planetary magnetic field strength, there should be more heating on planets subject to higher levels of irradiation \citep{Perna2012}, but these planets will also experience stronger drag on their winds and we may expect an anti-correlation between the amount of radius inflation and the offset of the hot spot from the substellar point \citep{Menou2012,RM12b}.  There is evidence for temperature inversions in the atmospheres of many hot Jupiters \citep[e.g.,][and references therein]{Madhu2010} and, while the presence of clouds or stratospheric absorbers will influence the pressure levels at which ohmic heating occurs \citep{Heng2012}, it is possible that the temperature inversions are in fact produced by ohmic heating \citep{Menou2012b}.  Finally, with future observing facilities we may be able to directly measure the upper atmosphere wind speeds on these planets and thereby constrain the strength of magnetic drag \citep{Kempton2012}.

Due to the complexity of this topic, most of the work so far has estimated ohmic heating rates from three-dimensional circulation models or analytic assumptions, without consistently including the feedback of magnetic drag and heating on the circulation pattern.  We have previously published models that attempted to estimate the effect of magnetic drag on hot Jupiter circulation by including a simple, approximate form for the drag \citep[][also used for Miller-Ricci Kempton \& Rauscher 2012]{Perna2010a,RM12b}, but here we improve upon that work in several important ways.  First, instead of estimating the drag strength, we now calculate it directly from local conditions (temperature, density) and update it with each timestep, which makes it strongly coupled to the dynamics.  This also means that the magnetic drag is no longer uniform at each pressure level, but instead varies by many orders of magnitude around the planet.  In addition, we include geometric effects due to an assumed aligned dipole field: only the zonal (east-west) component of the flow experiences drag and the strength of drag is also dependent on latitude (with zero drag at the equator).  Finally, we convert the kinetic energy lost through drag into heating from ohmic dissipation.  By including all of these effects, we present the first atmospheric circulation model with geometrically and energetically consistent magnetic drag and ohmic dissipation.

One obvious unknown in studying this topic is the strength of hot Jupiter magnetic fields.  We can use scaling laws to predict field strengths, but our knowledge is limited by the unknown complexity of planet interiors and an incomplete understanding of dynamo theory \citep[see reviews by][]{Christensen2010,Stevenson2010}.  Nevertheless, such scaling laws estimate hot Jupiter magnetic field strengths to be anywhere from a few to tens of Gauss \citep[see][and references therein]{Reiners2010}.  Of course it would be preferable to actually measure the magnetic field strength for any particular planet, but unfortunately this will most likely require the use of indirect methods in which the signature of the planet is buried in the stellar signal.  These include the measurement of changes in the stellar chromospheric or X-ray emission due to interaction between its and an orbiting planet's magnetic field (e.g., Shkolnik et al. 2005, 2008, Kashyap et al. 2008; although see Poppenhaeger et al. 2010, 2011, Miller et al. 2012), direct detection of radio emission from the planet \citep[which is likely too weak for current capabilities, see][and references therein]{Griessmeier2011}, and the signature of a planet's magnetic field in its influence on the rate and geometry of atmospheric evaporation \citep[e.g.][]{Yelle2004,Adams2011,Trammell2011} or by the field mediating the location of a shock between the planet and its host star's coronal plasma \citep{Vidotto2010,Vidotto2011}.

We describe our numerical model in Section~\ref{sec:code}, with particular attention to the new implementation of magnetic effects.  In Section~\ref{sec:models} we catalog the set of models presented in this paper, explaining our choices for the range of physical parameters used.  We begin the analysis of our results with a detailed look at the model of HD 209458b with $B=3$ G (\ref{sec:hd2b3}), followed by an examination of models with increasing magnetic field strengths (\ref{sec:bfield}), and a comparison between our models of HD 209458b and the cooler planet HD 189733b (\ref{sec:hd1}).  We then discuss the influence of numerical resolution on our results (\ref{sec:resolution}).  In Section~\ref{sec:obs} we examine the observable properties of our models and in Section~\ref{sec:radii} we discuss global ohmic heating rates, commenting upon how they could influence a planet's thermal evolution and its radius.  We conclude with a summary of our main findings in Section~\ref{sec:conc}.

\section{Our Numerical Model} \label{sec:code}

We have updated our General Circulation Model (GCM) to include the effects of geometrically and energetically consistent magnetic drag and heating.  Our GCM is adapted from the Intermediate General Circulation Model (IGCM) originally developed by \citet{Hoskins1975}, with a pseudo-spectral dynamical core that solves the primitive equations of meteorology.  
The vertical coordinate is $\sigma = P/P_s$, where $P$ is pressure and $P_s$ is the bottom boundary ``surface'' pressure.  No flow is allowed through either the top or bottom boundary, by imposing $\dot{\sigma}=0$ at $\sigma=0$ and 1.
We discuss our adaptation of the code to study synchronously rotating gas giants in \citet{MR09} and \citet{RM10}, and our implementation of a standard two-stream, double-gray radiative transfer scheme in \citet{RM12b}, hereafter RM12.  For our radiative transfer scheme, we separate the optical and infrared wavelengths, with the attenuation of the incident stellar (optical) flux controlled by a constant optical absorption coefficient, while the absorption and emission of thermal radiation at each level is set by a constant infrared absorption coefficient.

\subsection{Modeling the magnetic effects} \label{sec:magmodel}

Throughout the entire body of a gas giant planet there is a range of possible ionization levels and multiple processes responsible.  In the outer atmosphere, at pressures of nanobars, we expect photoionization due to the incident stellar UV radiation.  Very deep within the planet, at pressures of megabars, there should be phase transition to metallic hydrogen, with the associated increase in conductivity.  Here we focus on the range of pressures typically included in atmospheric circulation models, from $\sim$1 mbar to $\sim$100 bar.  In this region the temperatures on hot Jupiters may be high enough (depending on the stellar insolation) that there is sufficient thermal energy to ionize elements, especially trace alkali metals, due to their low first ionization potentials.  As we detail below, this should lead to weak ionization, where the ions are embedded in a mostly neutral flow.

We employ several simplifying assumptions in order to model complex magnetic processes without using a full magnetohydrodynamics (MHD) calculation.  In \citet{Perna2010a} we found that throughout the modeled atmosphere (from 1 mbar to 100 bar) the magnetic Reynolds numbers are generally much less than 1 and we compared the relative importance of the non-ideal MHD terms in the induction equation to determine that most of the atmosphere is within the purely resistive MHD regime, meaning that the Hall and ambipolar diffusion terms can be neglected in comparison to the Ohmic term, although there are localized regions where these assumptions can break down.  We additionally assume that the planet's magnetic field (presumably generated deep within the interior) has a dipole geometry and is aligned with the planet's rotation axis; it also remains unaltered by the weakly ionized flow in the atmosphere.

We only apply magnetic drag to the zonal (east-west) wind ($u$), leaving the meridional (north-south) wind unaltered, by adding a term $du/dt=-u/\tmag$ to the momentum equation.  In the case of an aligned dipole field, any drag on the meridional flow is not significant until near the poles, where the field becomes more radial.  As a simplification we do not vary the magnetic field strength along the planet surface---as it should for a true dipole---effectively ignoring the magnetic geometry at the poles.  We choose to use a constant field strength instead of more complex geometry in part because the detailed geometry of the actual planetary field is unknown.  However, the meridional flow can be fairly significant over the poles, especially high in the atmosphere.  Future work will be required to expand the magnetic formalism we use and to determine how large of an impact meridional drag could have on the circulation.

The magnetic timescale (\tmag) is calculated from the chosen magnetic field strength ($B$) and local conditions (density, $\rho$; temperature, $T$; and latitude $\phi$):
\begin{equation}
\tmag (B, \rho, T, \phi) = \frac{4\pi \rho \ \eta(\rho,T)}{B^2 \ | \sin \phi |}.  \label{eqn:tmag}
\end{equation}
\noindent (For a derivation of \tmag, see Perna et al. 2010a.)  Note the geometric dependence of \tmag; due to the assumptions of latitudinal currents and an aligned dipole field, $\vec{\jmath} \times \vec{B}=0$ at the equator ($\phi=0$) and there is no drag on the flow.  The resistivity ($\eta$) is calculated as in \citet{Menou2012}:
\begin{equation}
\eta = 230 \sqrt{T}/x_e \ \mathrm{cm^2 \ s^{-1}}  \label{eqn:eta}
\end{equation}
\noindent with the ionization fraction ($x_e \ll 1$) calculated from a form of the Saha equation that takes into account the first ionization potential ($\epsilon_i$) of all elements from hydrogen to nickel,\footnote{By using the first 28 elements, rather than just potassium \citep[as in][]{Perna2010a}, the resistivities and timescales are decreased by a factor of $\sim$2, for the conditions of interest here.} assuming solar abundance for each element ($a_i$):
\begin{eqnarray}
n_i &=& \left(\frac{a_i}{a_{H}}\right) n \\
\frac{x_{e,i}^2}{1-x_{e,i}^2} &=& \frac{1}{n_i k T} \left(\frac{2 \pi m_e}{h^2}\right)^{3/2} (kT)^{5/2} \exp(-\epsilon_i/k T) \\
x_e &=& \sum_{i=1,28} \left(\frac{n_i}{n}\right) x_{e,i}
\end{eqnarray}
\noindent where $n$ is the number density ($=\rho/\mu$, with $\mu$ being the mean molecular weight).  The local resistivities and magnetic timescales throughout the entire atmosphere are updated at each timestep in the simulation, making the magnetic effects strongly coupled with the atmospheric circulation.

We assume that all of the kinetic energy lost via magnetic drag is returned to the atmosphere as localized ohmic heating.  The power from ohmic dissipation is an extra term added to the energy equation and we calculate it as \citep{Liu2008,Perna2010a,Menou2012}:
\begin{equation}
 \left (c_p \frac{dT}{dt}\right)_{\mathrm{ohm}} = \frac{1}{\rho} \frac{4 \pi \eta}{c^2} j^2
	= u^2 \frac{B^2 | \sin \phi |}{4 \pi \eta \rho} = \frac{u^2}{\tmag},  \label{eqn:mheat}
\end{equation}
\noindent which is energetically consistent with our form for the drag ($du/dt=-u/\tmag$).

\begin{figure}[ht!]
\begin{center}
\includegraphics[width=0.45\textwidth]{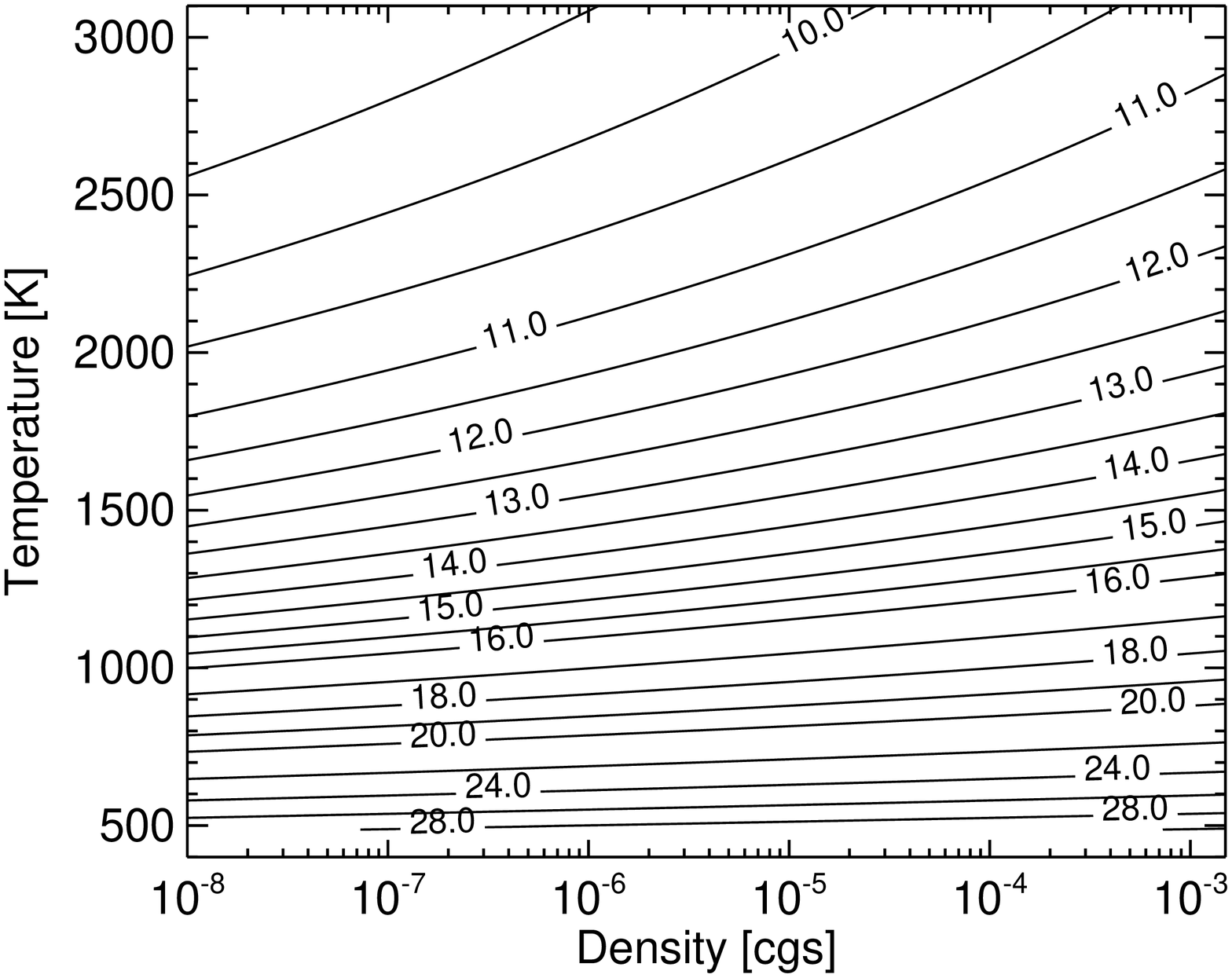}
\includegraphics[width=0.45\textwidth]{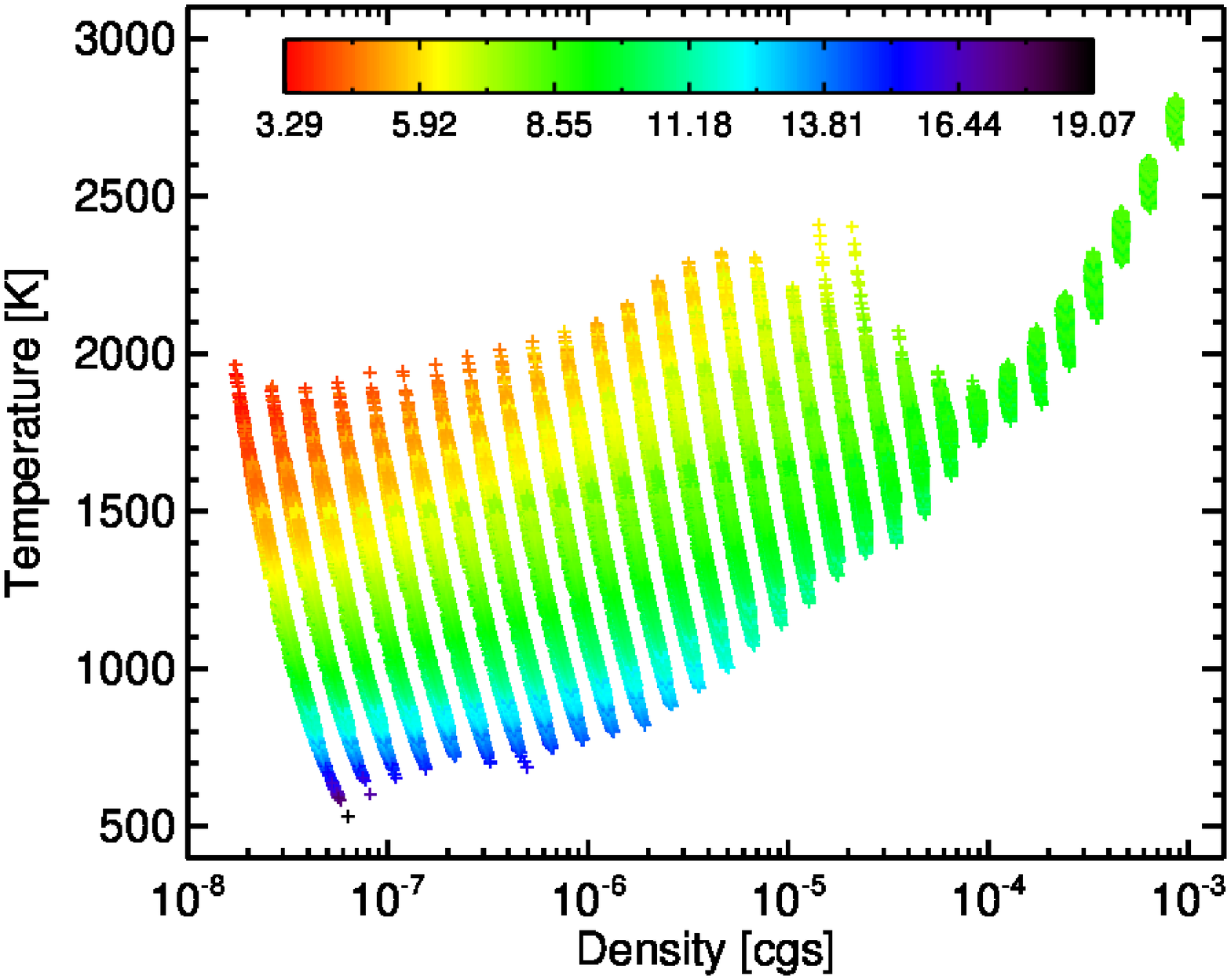}
\end{center}
\caption{Left: the electrical resistivity of the atmosphere, $\eta$, as a function of the local temperature and density, over the range of values expected in hot Jupiter atmospheres.  The contour levels give $\log_{10}(\eta)$ in cm$^2$ s$^{-1}$.  Right: the timescale for magnetic drag and heating, \tmag, for our model of HD~209458b with $B=3$ G.  The color of each point gives $\log_{10}(\tmag)$ in seconds.  Each group of points (roughly vertically aligned) corresponds to one of the discrete pressure levels in the simulation.  The variations in \tmag~for a given density and temperature are due to the geometric dependence of \tmag~(see Equation~\ref{eqn:tmag}).} \label{fig:eta}
\end{figure}

In Figure~\ref{fig:eta} we plot the electrical resistivity ($\eta$) as a function of temperature and density, over the range of values expected for hot Jupiter atmospheres.  Although only weakly dependent on density, the resistivity is a strong function of temperature.  A second plot in the same figure shows an example of how the magnetic timescale (\tmag) varies throughout a planet's atmosphere.  Timescales are generally shorter deeper in the atmosphere, but the strongest variation---by many orders of magnitude---is in the upper levels, between the hot dayside and the cold nightside.  Figure~\ref{fig:eta} also shows the geometric dependence of \tmag, resulting in differing timescales at the same values of the local density and temperature.

It is important to note that we are applying a formalism derived assuming axisymmetry \citep{Liu2008} to a non-axisymmetric problem.  Figure~\ref{fig:eta} shows the huge difference in resistivities between the day and night sides of the planet.  In addition, there are regions of the atmosphere (usually at low pressure) where the circulation is not dominated by a zonal jet (or jets), meaning that there is a significant non-axisymmetric component.  As we will see, this can become even more of an issue once the magnetic effects are included in the models (for example, see the flow patterns at low pressure in Figures~\ref{fig:r611} and~\ref{fig:r602}).  While the use of this formalism may be justified in first approximation, we would (as always) benefit from a fuller, non-axisymmetric theory.

In order to maintain numerical stability we prevent extremely strong drag or heating by setting a minimum magnetic timescale, \tmm, which is used instead of the timescale calculated from Equation~\ref{eqn:tmag} if it is greater than \tmag.  In the models presented here we set the minimum magnetic timescale equal to our hyperdissipation\footnote{Hyperdissipation is a common technique used to prevent the build up of numerical noise on the smallest resolved scales.} timescale ($=0.005$ in units of the planet's rotation period, \Prot).  For comparison, the timestep used in our models is $\approx0.0002$ \Prot.  Among the range of parameters we tested, it was only the model with strongest magnetic field strength ($B=30$ G) and the hottest atmosphere (HD 209458b) where our choice of $\tmm=0.005$ limited the full strength of the magnetic effects (and that model failed to run to completion, see Section~\ref{sec:bfield}).

We initialized each model with the atmosphere at rest (no winds).  The initial temperature structure was set to be uniform at each pressure level, using a temperature-pressure profile from \citet{Guillot2010} with parameters matching those used for our radiative transfer, and with the averaging parameter set to $f=0.375$, between a global average ($f=0.25$) and a dayside average ($f=0.5$), in order to minimize the time needed for the atmosphere to equilibrate.  (A global average was not chosen because it would have led to very strong heating in the substellar regions.)  The atmosphere was then allowed to heat and cool according to our radiative transfer scheme, accompanied by the acceleration of winds.  After 10 orbital periods we introduced the magnetic effects of drag and heating, linearly increasing from zero to reach their full values at 20 orbital periods.  We tested using 2/4 \Prot~or 50/100 \Prot~instead of 10/20 \Prot~for the implementation of magnetic effects.  By waiting to apply the magnetic effects until later in the run, the winds were able to accelerate to faster values early in the simulation, but by 1000 \Prot~we found no significant difference between the models.

\subsection{Our set of models} \label{sec:models}

We have tested the effects of magnetic drag and heating over a range of physical and numerical parameters.  The default resolution we used was T31L30, the horizontal spectral resolution corresponding to $\sim$4\degrees~and the 30 vertical levels logarithmically spaced in pressure, from 1 mbar to 100 bar.  These are the same parameters we used in the models without magnetic effects from RM12 and we have also chosen the same hyperdissipation strength ($\nabla^8$, $\tau=0.005$ \Prot) and the same length for the runs (out to 2000 \Prot $=P_{\mathrm{orb}}$).  In Section~\ref{sec:resolution} we will discuss our tests of numerical resolution.

We ran models for planetary magnetic field strengths of $B=0$, 3, 10, and 30 G.  As discussed in Section~\ref{sec:intro}, it is not clear what field strength to expect for Jupiter-mass planets on orbital periods of a few days, so our values span a range from weak to strong fields.  We used planetary parameters meant to represent two well known hot Jupiters: HD 189733b and HD 209458b, as shown in Table~\ref{tab:params}.  In addition to being the two hot Jupiters with the most measurements, the difference in equilibrium temperature between these two planets ($\Delta T_{\mathrm{eq}} \approx 300$ K) means that we can test two levels of thermal ionization, with a difference of $\sim$2 orders of magnitude in electrical resistivities.  Although the physics of radiative transfer is highly detailed and differences between planets, particularly in composition, can lead to variation in their radiative properties, here we choose a more straightforward approach and use the same optical and infrared absorption coefficients in for both planets.  The hotter atmosphere of our HD 209458b model is solely the result of a higher incident stellar flux than for the HD 189733b model.  Note that we use different planetary radii, surface gravities, and rotation rates for these two planets, as appropriate, but these differences are secondary compared to the disparate levels of stellar heating.

\begin{deluxetable}{lccc}
\tablewidth{0pt}
\tablecaption{Model parameters used}
\tablehead{
\colhead{Parameter}  &  \colhead{HD 189733b} & \colhead{HD 209458b}  & \colhead{Units}
}
\startdata
Radius of the planet, $R_p$ 	& $8\times 10^7$ 		& $1\times 10^8$ 		& m \\
Gravitational acceleration, $g$ & 22					& 8 					& m s$^{-2}$ \\
Rotation rate, $\Omega$ 		& $3.3 \times 10^{-5}$ 	& $2.1 \times 10^{-5}$ 	& s$^{-1}$ \\
\ \ \ Corresponding period, $P_{\mathrm{rot}} = 2 \pi /\Omega$ & 2.2 & 3.3 	& day$_{\oplus}$ \\
Equilibrium temperature\tablenotemark{a}, $T_{\mathrm{eq}}$  & 1200  &  1500 &  K \\
Incident flux at substellar point, $F_{\downarrow \mathrm{vis}, \mathrm{irr}}$ & $4.74 \times 10^5$ & $1.06 \times 10^6$  & W m$^{-2}$  \\
\ \ \ Corresponding temperature, \Tirr & 1700 & 2078 & K \\
\hline
Internal heat flux, $F_{\uparrow \mathrm{IR}, \mathrm{int}}$ & \multicolumn{2}{c}{3500} & W m$^{-2}$ \\
\ \ \ Corresponding temperature, \Tint & \multicolumn{2}{c}{500} & K \\
Optical absorption coefficient, \kvis & \multicolumn{2}{c}{$4 \times 10^{-3}$} & cm$^2$ g$^{-1}$ \\
Infrared absorption coefficient, $\kappa_{\mathrm{IR},0}$  & \multicolumn{2}{c}{$1 \times 10^{-2}$}  & cm$^2$ g$^{-1}$ \\
Infrared absorption powerlaw index, $\alpha$ & \multicolumn{2}{c}{0} & -- \\
Specific gas constant, $\mathcal{R}$ & \multicolumn{2}{c}{3523} & J kg$^{-1}$ K$^{-1}$ \\
Ratio of gas constant to heat capacity, $\mathcal{R}/c_P$ & \multicolumn{2}{c}{0.286} & -- \\
\enddata
\label{tab:params}
\tablenotetext{a}{assuming an albedo of zero}
\end{deluxetable}

The level of thermal ionization in a planet's atmosphere is dependent both on temperature and on the abundance of elements with low ionization potentials (particularly alkali metals).  While the atmosphere of HD 189733b should be cooler than that of HD 209458b, there is no strong constraint on the metallicities of these planets.  The uncertainty in atmospheric composition factors into the resistivity and may be somewhat balanced by our uncertainty in the magnetic field strength, since these parameters together control the strength of the magnetic effects (Equation~\ref{eqn:tmag}).  In order to test whether enhanced metallicity and a strong magnetic field could compensate for the cooler temperatures on HD 189733b, we ran a model of this planet with $B=30$ G and metal abundances increased $3\times$ above solar, a value arbitrarily chosen to mimic Jupiter's composition \citep{Wong2004} and in lieu of a consensus on the measured metallicity for this planet.

The physical effects that result from increased atmospheric metallicity are: 1) an increased abundance of ions from thermal ionization, 2) an increase in the mean molecular weight (MMW), and 3) a change in opacities throughout the atmosphere.  While the slight increase in MMW could have a small effect on the atmospheric dynamics, changing the opacities is known to have a large effect \citep{DobbsDixon2008,Showman2009,Lewis2010}, since this controls the structure of radiative heating throughout the atmosphere.  However, here we are only going to adjust the elemental abundances in the ionization calculation, without changing the other parameters of the model, in order to isolate the effect of increased ionization.  Since we are using two constant absorption coefficients for our radiative transfer, and these were chosen in order to roughly match expected temperature-pressure profiles from 1D models, there is no clear prescription for how we would change our opacities to match the change in metallicity.  An interesting avenue for future work will be atmospheric circulation models that investigate the dual influence of metallicity on radiative heating and on magnetic effects.

Finally, we test the importance of the internal heat flux (characterized by \Tint) on the amount of ohmic heating in the atmosphere.  We do not expect our choice for \Tint~to affect the atmospheric circulation in the upper atmosphere.  This is to be expected, given that the stellar irradiation is more than two orders of magnitude stronger than any reasonable value for the heat flux from the interior and therefore is the dominant driver of the circulation.  Our non-magnetic models in RM12 confirm that the circulation near the infrared photosphere is insensitive to the value chosen for \Tint.  However, whether ohmic heating could lead to radius inflation depends not just on the strength of the heating, but also on the depth at which it occurs, with deeper heating more likely to influence the global structure of the planet \citep{Guillot2002}.  The deep atmosphere will be hotter for higher values of \Tint~and we expect that this should result in more ohmic heating.  Although a full study could be made examining a range of \Tint, we will leave that for future work and here simply compare $B=3$ G models of HD 209458b with $\Tint=500$ K (our default) and $\Tint=100$ K.

\section{Results} \label{sec:results}

We begin the presentation of our results with a detailed look at the model of HD 209458b with $B=3$ G in order to demonstrate how magnetic effects function in hot Jupiter atmospheric circulation.  Aside from the inclusion of magnetic drag and heating, this model is identical in all parameters to one presented in RM12, allowing for a clean comparison against a non-magnetic model.

\subsection{HD 209458b with $B=3$ G} \label{sec:hd2b3}

The first difference between this model and the non-magnetic version in RM12 is in the development of the flow, starting from an initial condition with no winds.  In the absence of magnetic drag the atmosphere accelerates, first in the directly forced region of the atmosphere (above $\sim$1 bar) and then, through the vertical transfer of momentum, in the levels below the photosphere.  Eventually the winds reach their peak speeds and the atmosphere is in a quasi-steady state.  We find that the presence of magnetic drag alters this pattern.  While there is still an initial acceleration of winds in the upper and lower levels, this is quickly followed by the response of the magnetic drag, decelerating the winds.  This initial ramp-up/ramp-down period lasts $\sim$1200 \Prot~for the HD~209458b $B=3$ G model, after which follows repeated acceleration and deceleration of the winds, with no strict periodicity, but a rough timescale of $\sim$100 \Prot~(Figure~\ref{fig:diag_uz} shows this behavior, in the zonally averaged flow at the equator).  We find that the entire atmosphere gains and loses kinetic energy (and total enthalpy) with these variations, although by the end of the 2000 \Prot~simulation the variations in total energies are less than $\sim$1\%.

\begin{figure}[ht!]
\begin{center}
\includegraphics[width=0.4\textwidth]{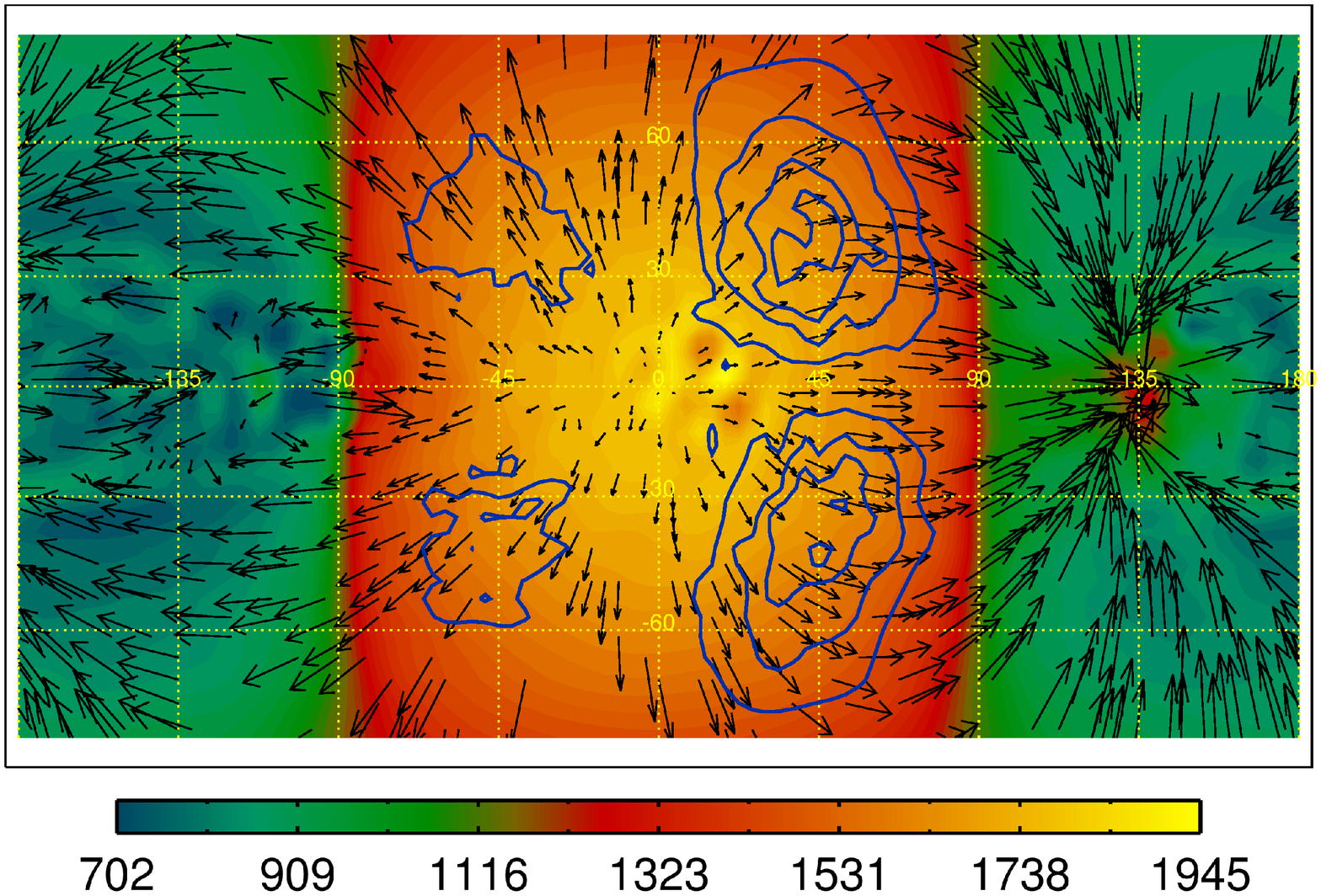}
\includegraphics[width=0.4\textwidth]{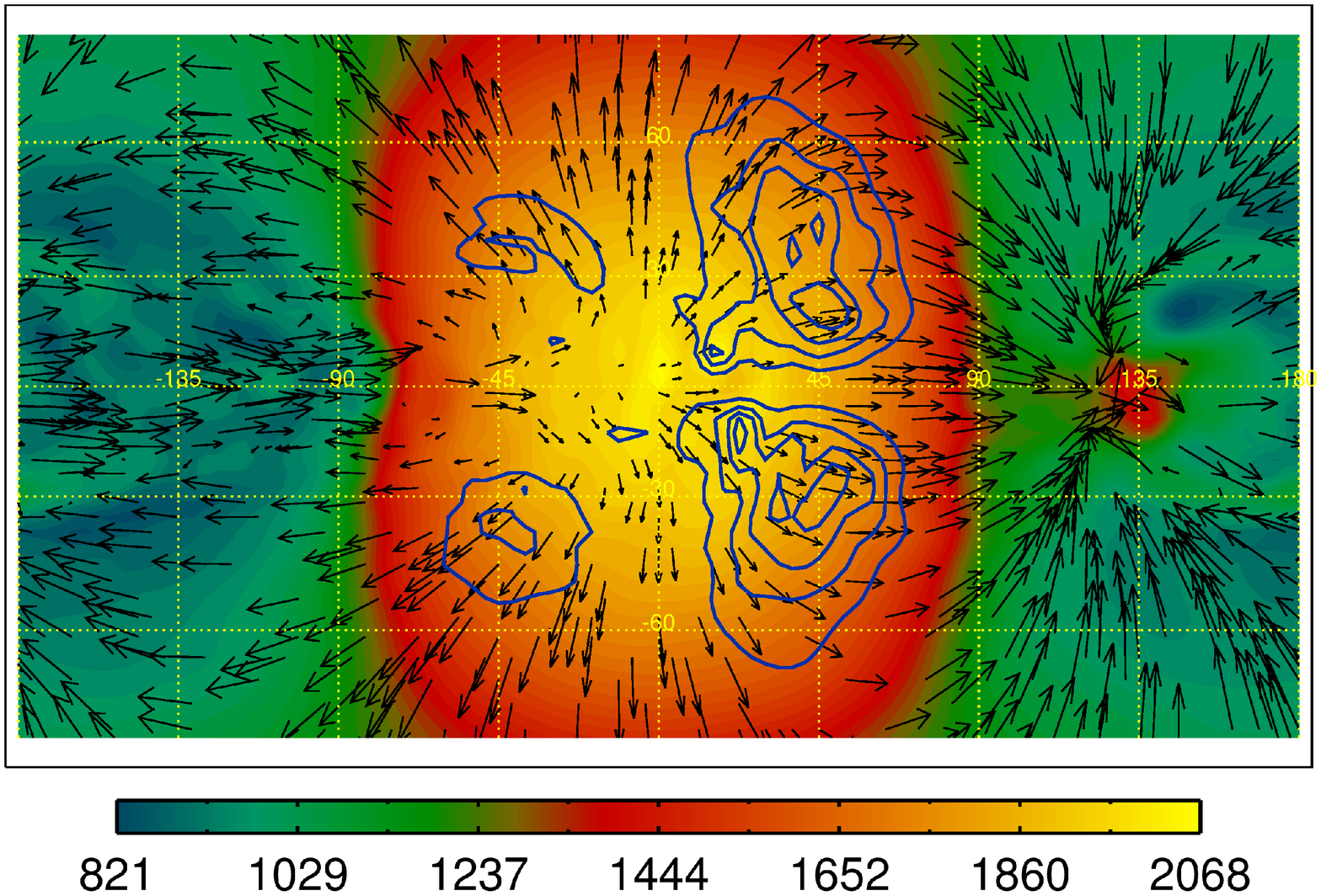}
\includegraphics[width=0.4\textwidth]{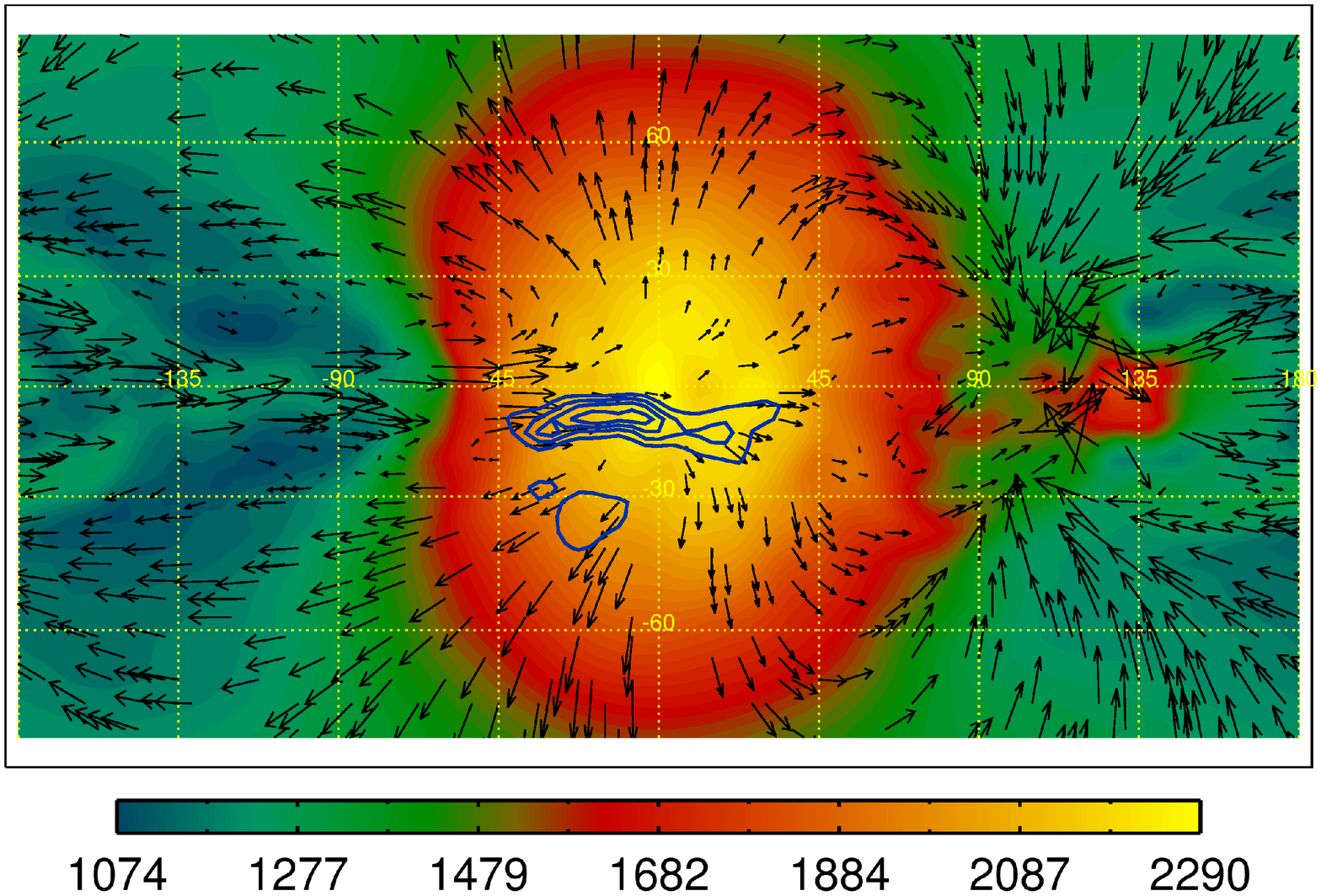}
\includegraphics[width=0.4\textwidth]{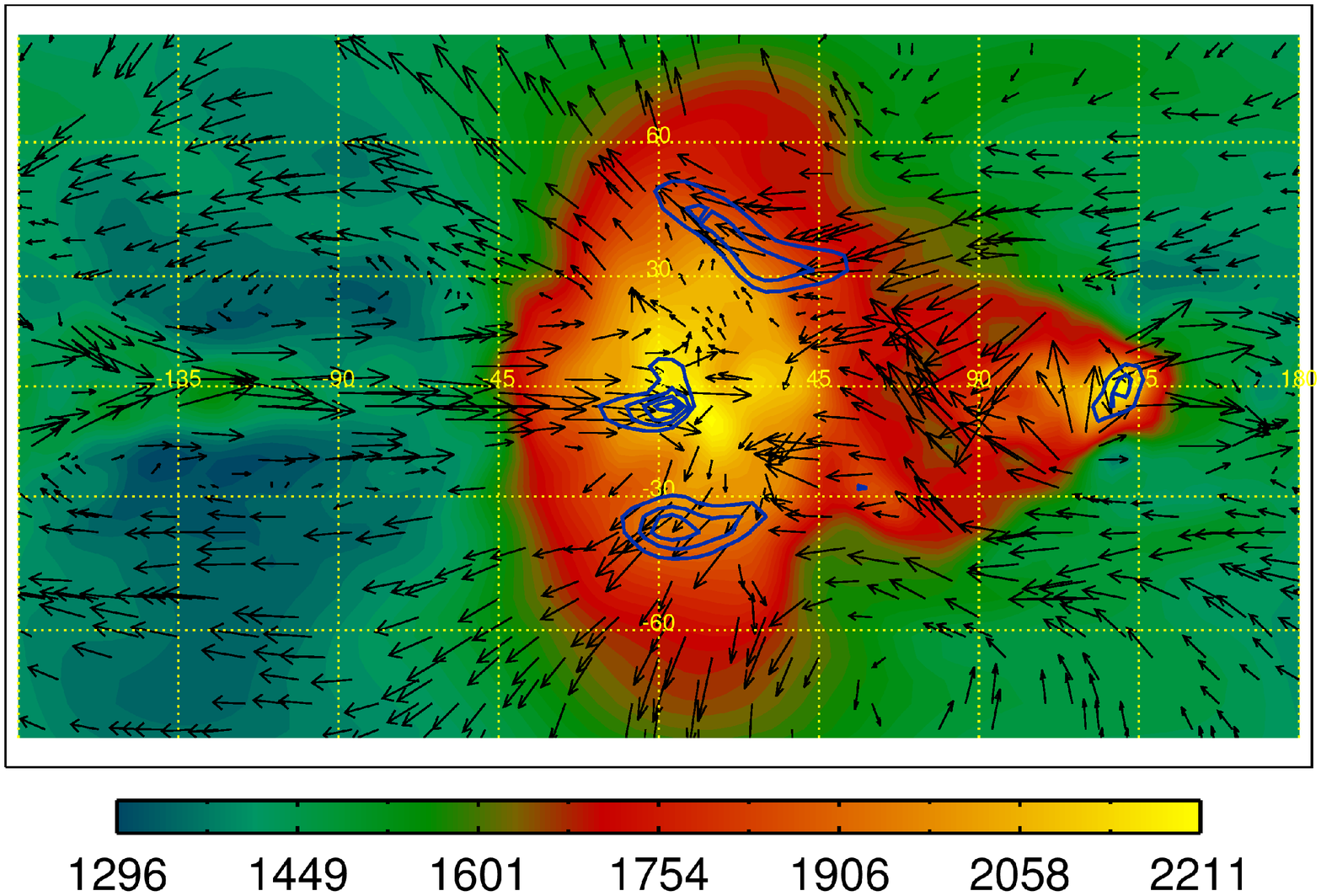}
\includegraphics[width=0.4\textwidth]{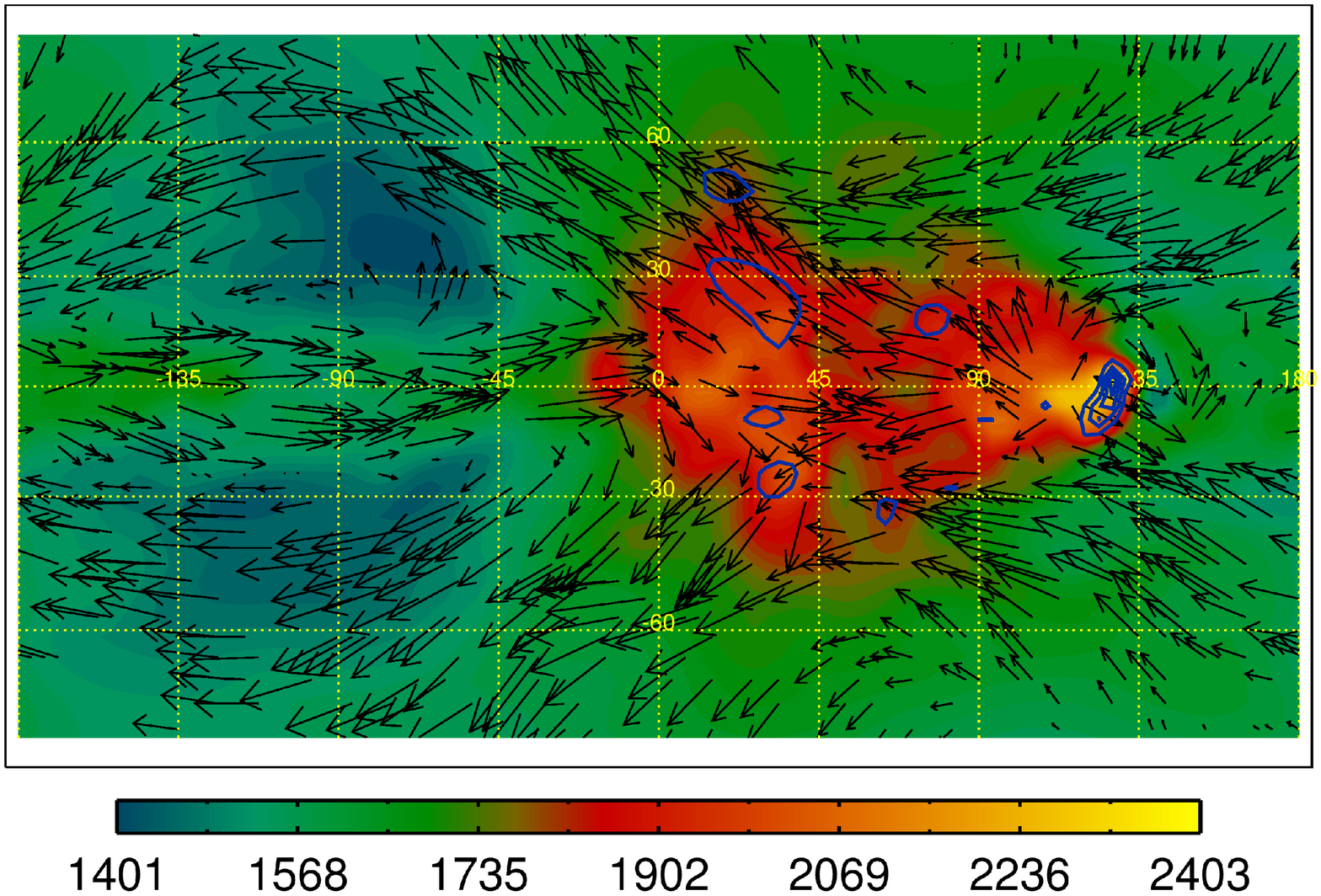}
\includegraphics[width=0.4\textwidth]{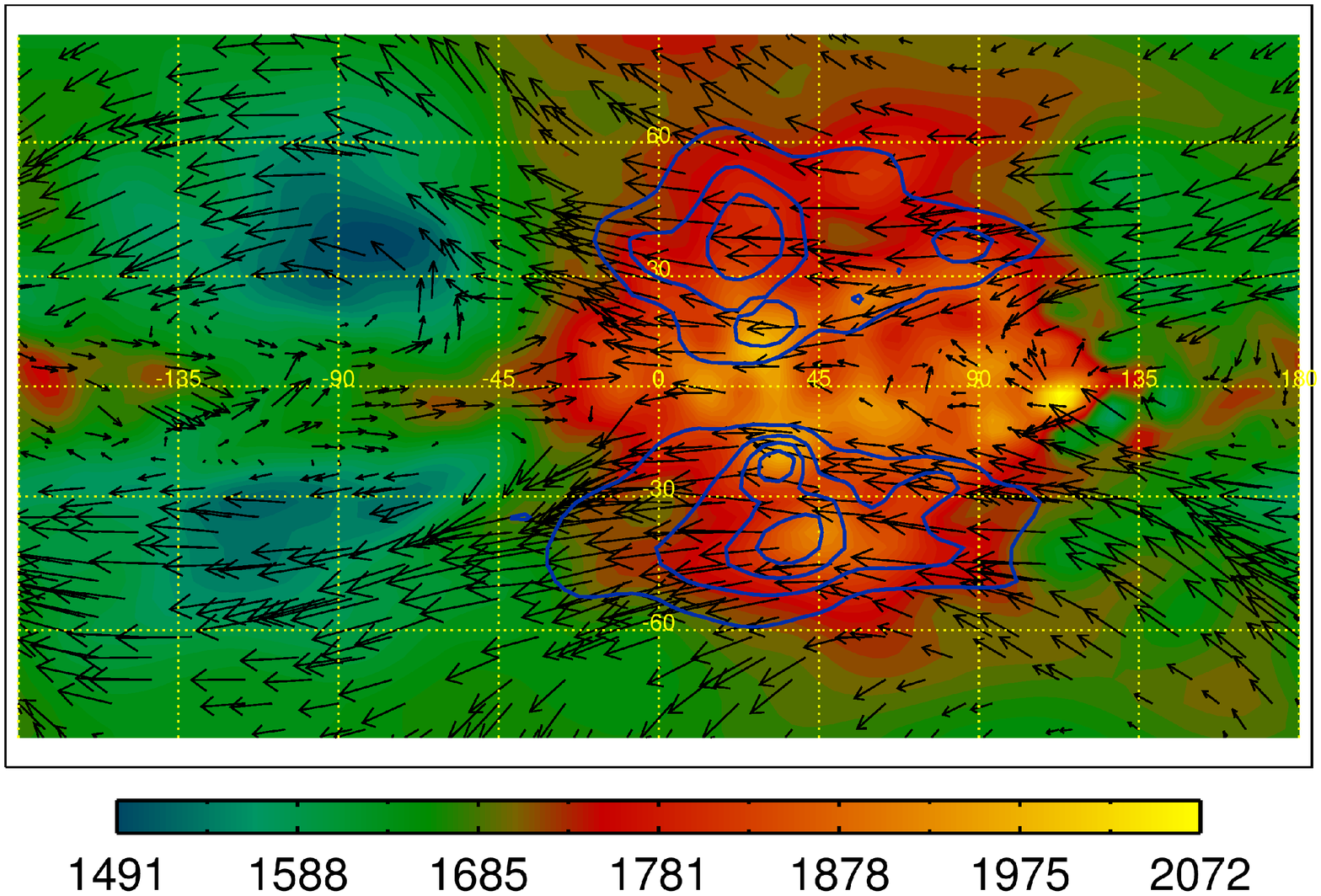}
\end{center}
\caption{Ohmic heating (shown as blue contours at 20, 40, 60, and 80\% of the peak value, and coincident with the magnetic drag) for the model of HD 209458b with $B=3$ G, plotted with winds (arrows) and temperatures (color scale, in K).  These horizontal slices through the atmosphere (in cylindrical projection, centered on the substellar point) are at pressure levels of 8 mbar (top left), 56 mbar (near the infrared photosphere, top right), 260 mbar (middle left), 800 mbar (middle right), 2 bar (bottom left), and 3 bar (bottom right), with peak wind speeds of 8.1, 5.9, 4.6, 2.5, 1.2, and 1.3 \kms, and peak ohmic heating rates of 22, 3.4, 1.6, 0.10, 0.014, and 0.0039 \Wkg, respectively.} \label{fig:r611}
\end{figure}

Although the presence of magnetic drag slows average wind speeds by $\sim$1 \kms~compared to the non-magnetic model (see Figure~\ref{fig:pprofs}), the flow in the upper atmosphere is still supersonic ($c_s \approx 3$ \kms).
In addition to slowing the winds, the presence of magnetic drag also disrupts the development of the strong super-rotating equatorial jet seen throughout the atmospheres of most hot Jupiter models (see Figure~\ref{fig:drag_uz}).
Even though there is no drag along the equator itself (see Equation~\ref{eqn:tmag}), the equatorial jet is inhibited because the drag disrupts the mechanism of slanted, up-gradient transport of angular momentum identified by \citet{Showman2011}.  Those authors recognized that eastward propagation of Kelvin waves, together with westward propagation of equatorially trapped Rossby waves, sets up a slanted flow pattern on hot Jupiters that channels eastward momentum to the equator.  \citet{Showman2011} found that very strong drag can limit the differential zonal propagation of these waves and prevent the slanted transport of angular momentum and the development of equatorial super-rotation.  The drag present in our models differs from the form used in their analysis in that it only acts on the zonal wind and is dependent on local conditions, but should similarly be able to disrupt the jet-pumping mechanism.
This disruption of the standard hot Jupiter circulation pattern has important consequences for the temperature structure of the atmosphere, as seen in Figure~\ref{fig:r611}, where we show horizontal slices at different depths in the atmosphere.  (We also plot the specific ohmic heating rates in this figure, calculated from the local conditions as per Equations~\ref{eqn:tmag} through~\ref{eqn:mheat}.)

The temperature structure in this model is more strongly tied to a hot day/cold night pattern than it is in the non-magnetic model.  Whereas in the non-magnetic model day-night temperature differences decrease with increasing pressure, and are negligible below $\sim$1 bar (see Figure 2 of RM12), in this magnetic model temperature contrasts of 1000 K are maintained down to at least 1 bar, only becoming homogenized below $\sim$4 bar.  The simplest explanation for this is that the slower wind speeds mean that gas heated on the day side has more time to cool before reaching the night side.  This is valid in the highest regions of the atmosphere, where we see the same type of substellar-to-antistellar (SSAS) flow as in the non-magnetic model; however, slightly different explanations are required at deeper levels.  If we compare the flow at the infrared photospheres of this and the non-magnetic model (see Figure 5 of RM12), we can see that in this magnetic model the hottest region of the atmosphere remains at the substellar point, instead of being advected eastward by tens of degrees in longitude.  At these pressures the difference in temperature structures is not just the result of slower winds, but also the lack of a strong eastward equatorial jet.  Instead, we find a flow that is still mostly SSAS with a convergent feature at $\sim$135\degrees~E, as at higher levels.

This convergent feature,\footnote{See Section 4.3 of \citet{RM10} for a discussion of this feature, found in many circulation models, including those from other groups \citep[e.g.,][]{Showman2009,DobbsDixon2012}.} seen across multiple levels, is a feature in common with the non-magnetic model.  The horizontal convergence leads to downward flow (by the continuity equation), which should result in adiabatic heating as the gas is pushed to higher pressures.  In the non-magnetic model we do see that at deeper pressure levels there is a distinct local hot spot associated with this feature, which is superimposed on the hot regions advected eastward from the day side.  In the model presented here, however, the hot spot remains an independent feature and is in fact enhanced by ohmic heating, as is especially apparent at the 2 bar level shown in Figure~\ref{fig:r611}.  Although this leads to an effective eastward shift in the hottest regions of the atmosphere, note that this is a distinctly different phenomenon from eastward \emph{advection} due to an equatorial jet.

Finally, although the 135\degrees~hot spot is an exception, the temperature structure in this model is influenced more strongly by the presence of magnetic drag, than by the (coupled) effect of ohmic heating.  When we compare rates of magnetic and radiative heating throughout the atmosphere, we find that almost everywhere the radiative heating dominates.  Deep in the atmosphere the ohmic heating becomes a non-negligible fraction of the radiative heating, but we will delay discussion of this point until our section about global heating rates (\ref{sec:radii}).


\subsection{Changes in the circulation due to increasing magnetic field strength} \label{sec:bfield}

We do not know the strength of hot Jupiter magnetic fields.  In our previous work on this topic we considered models of HD 209458b with magnetic field strengths of $B=3$, 10, and 30~G \citep[][RM12]{Perna2010a,Perna2010b} and we have used those same values here in order to facilitate comparison.  We have improved upon that previous work by including ohmic heating in addition to drag, by applying drag only to the zonal component of the wind, and by continuously calculating the strength of the magnetic effects based on changing local conditions, rather than just using a single value of \tmag~for each pressure level.  Using this updated code, we find that we are unable to successfully run models of HD 209458b with $B\geq 20$ G, for reasons we will discuss below.  To begin, we will compare our models of HD 209458b with $B=0$, 3, and 10 G.

\begin{figure}[ht!]
\begin{center}
\includegraphics[width=0.325\textwidth]{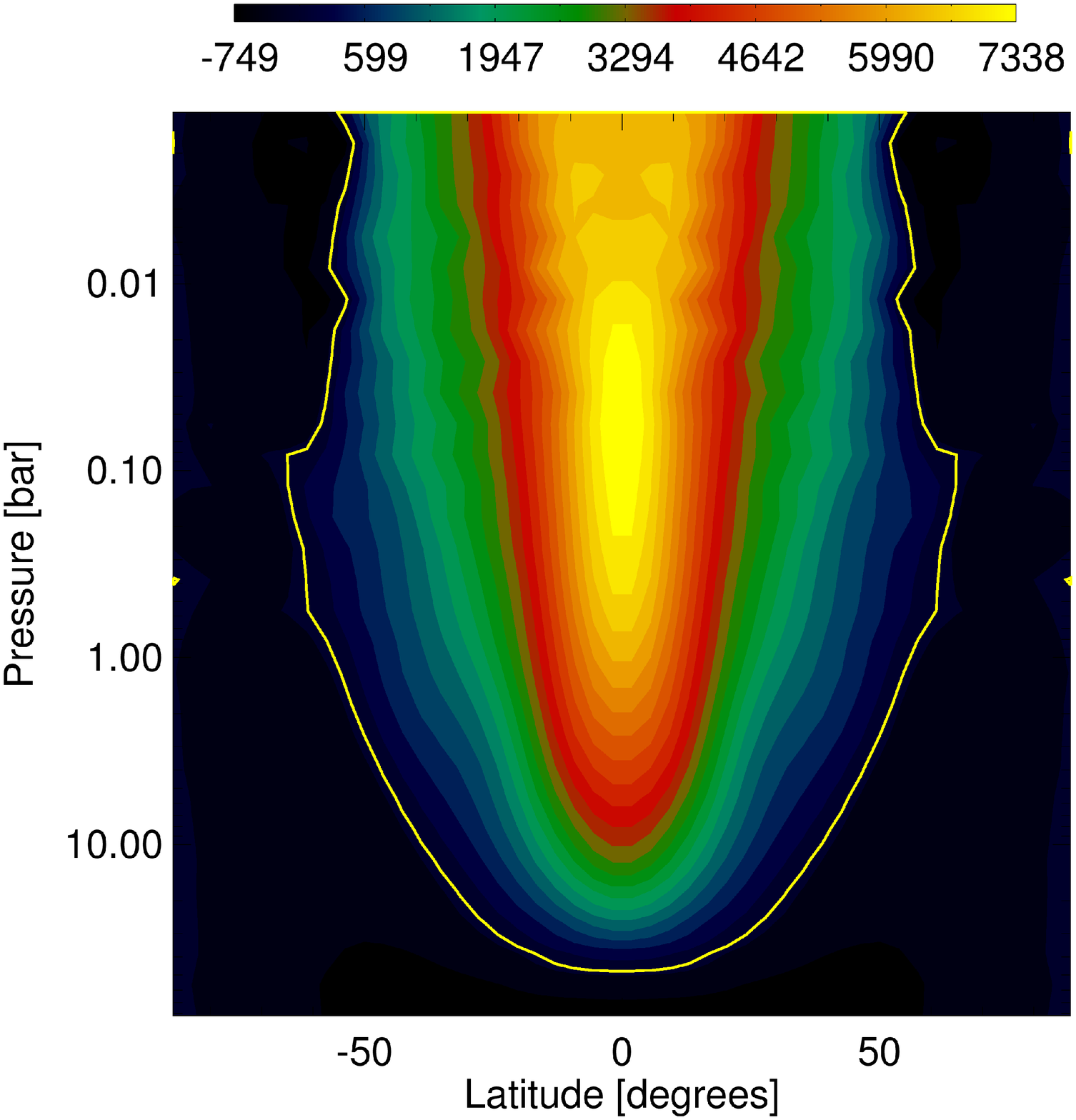}
\includegraphics[width=0.325\textwidth]{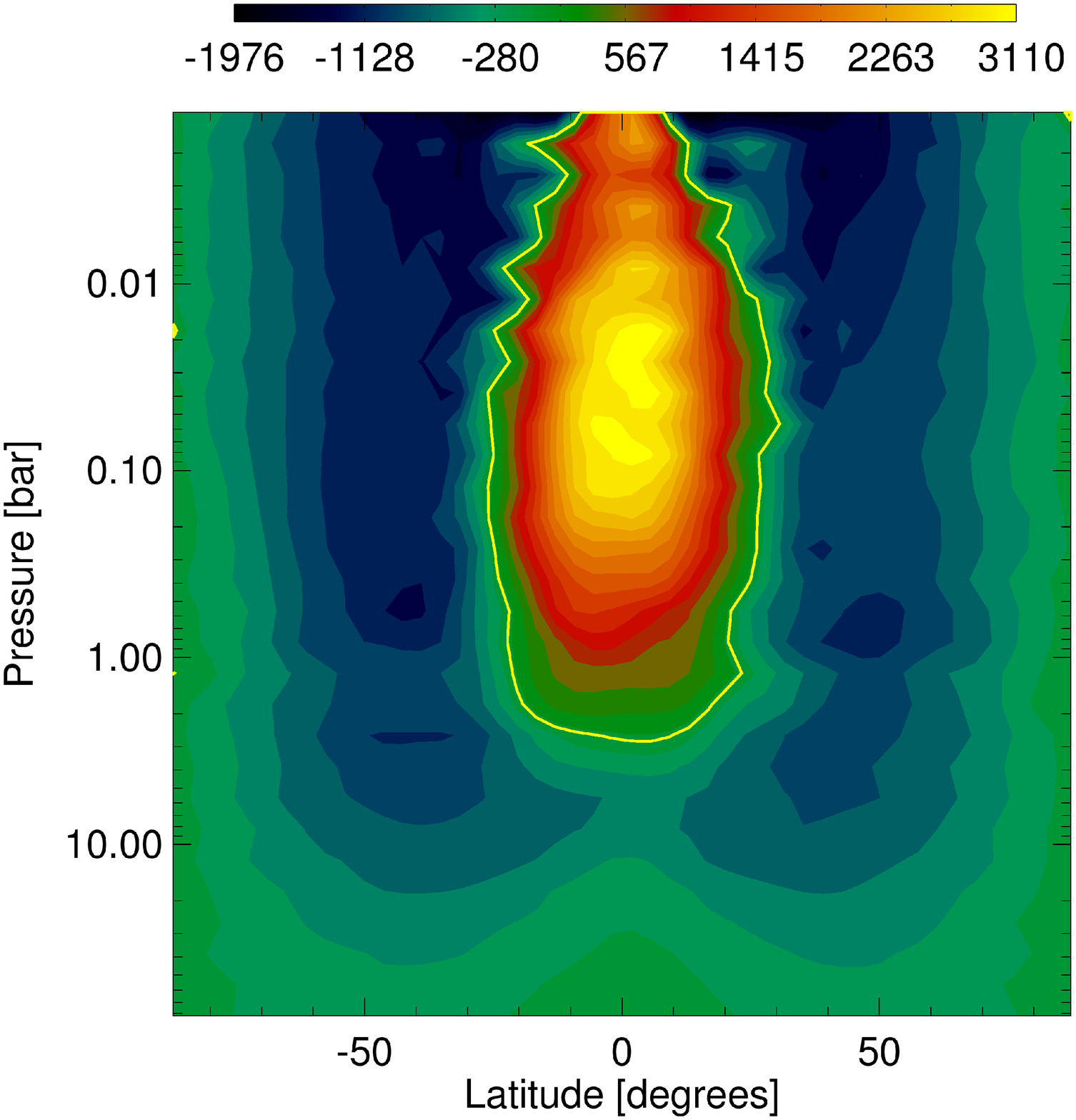}
\includegraphics[width=0.325\textwidth]{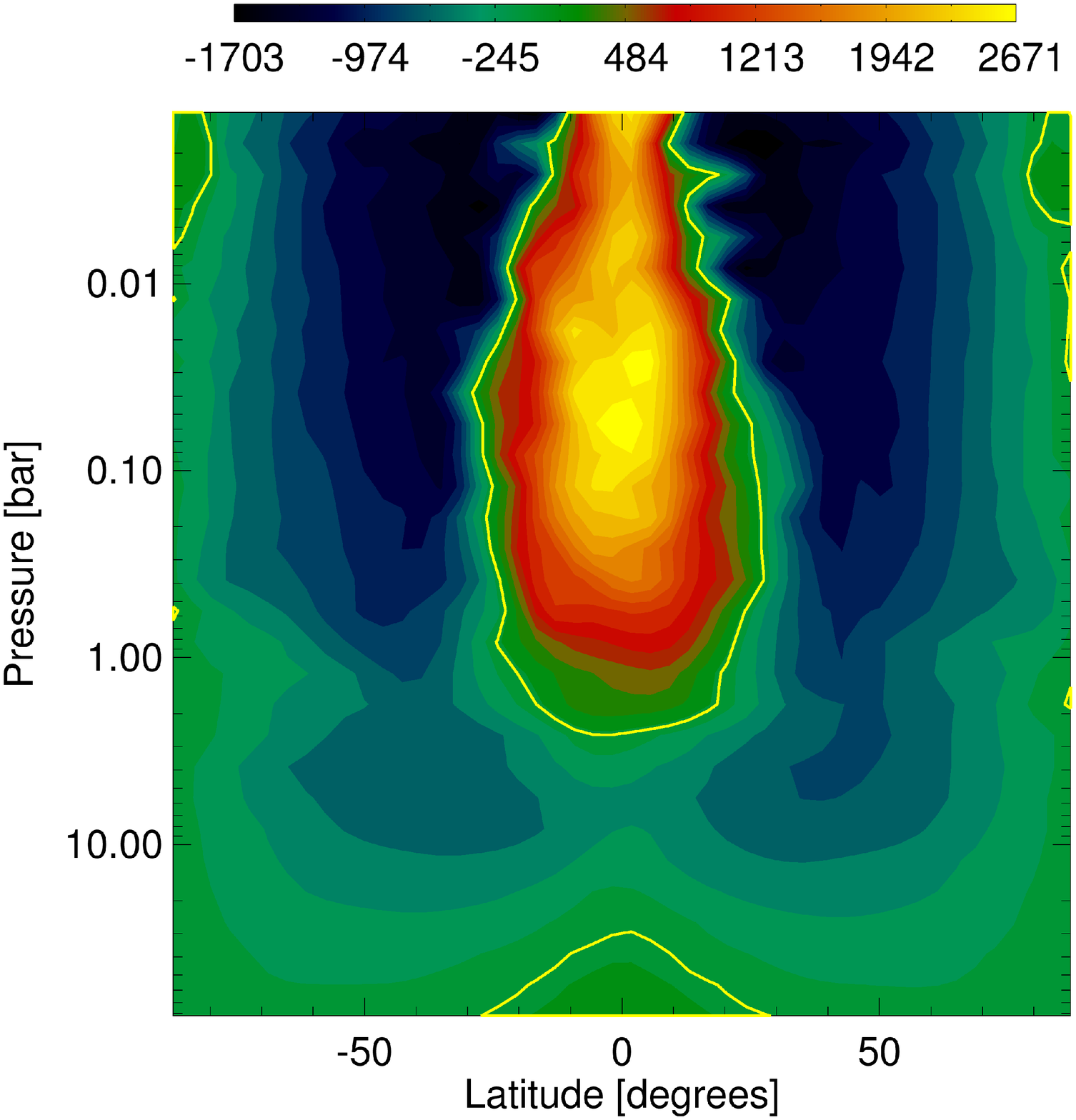}
\end{center}
\caption{Plots showing the zonal average of the zonal (east-west) wind (in \ms), as a function of latitude and pressure, for our models of HD 209458b with $B=0$ G (left), $B=3$ G (middle), and $B=10$ G (right).  The yellow line separates eastward (positive) from westward (negative) flow.  These are from snapshots of each model at 2000 \Prot.} \label{fig:drag_uz}
\end{figure}

In our previous models we saw a gradual change in circulation as $B$ was increased from 0 to 30 G \citep[][RM12]{Perna2010a}, but now with our more realistic scheme we see a sharp change in the circulation from the no-drag to the $B=3$ G model and then a more subtle change as the field strength is increased from $B=3$ to 10 G, as shown in the zonal wind profiles plotted in Figure~\ref{fig:drag_uz}.  The $B=10$ G model has similar circulation patterns as the $B=3$ G model, although with slightly slower wind speeds (but still supersonic) and with more of a departure from hemispheric symmetry (this can also be seen in Figure~\ref{fig:pprofs}), primarily due to an increased perturbation of the equatorial jet, causing it to meander farther to the north and south.

These same trends continue to higher magnetic field strengths, as seen in the developing flow of the $B=30$ G model, before it becomes numerically unstable and crashes.  Even with 1.5$\times$ more timesteps per \Prot, compared to the other models, this model would only run to 271 \Prot~before crashing.  (We also tested using $B=20$ G instead, but that model was likewise unable to run for the full 2000 \Prot.)
We have not been able to clearly identify the source of numerical instability in this model, although we speculate that it may be related to intense heating, as described below.
Although this uncompleted run cannot be used for a reliable picture of the circulation at $B=30$ G, it is nevertheless informative to examine the atmosphere before the crash in order to understand the limitations of the model.

\begin{figure}[ht!]
\begin{center}
\includegraphics[width=0.45\textwidth]{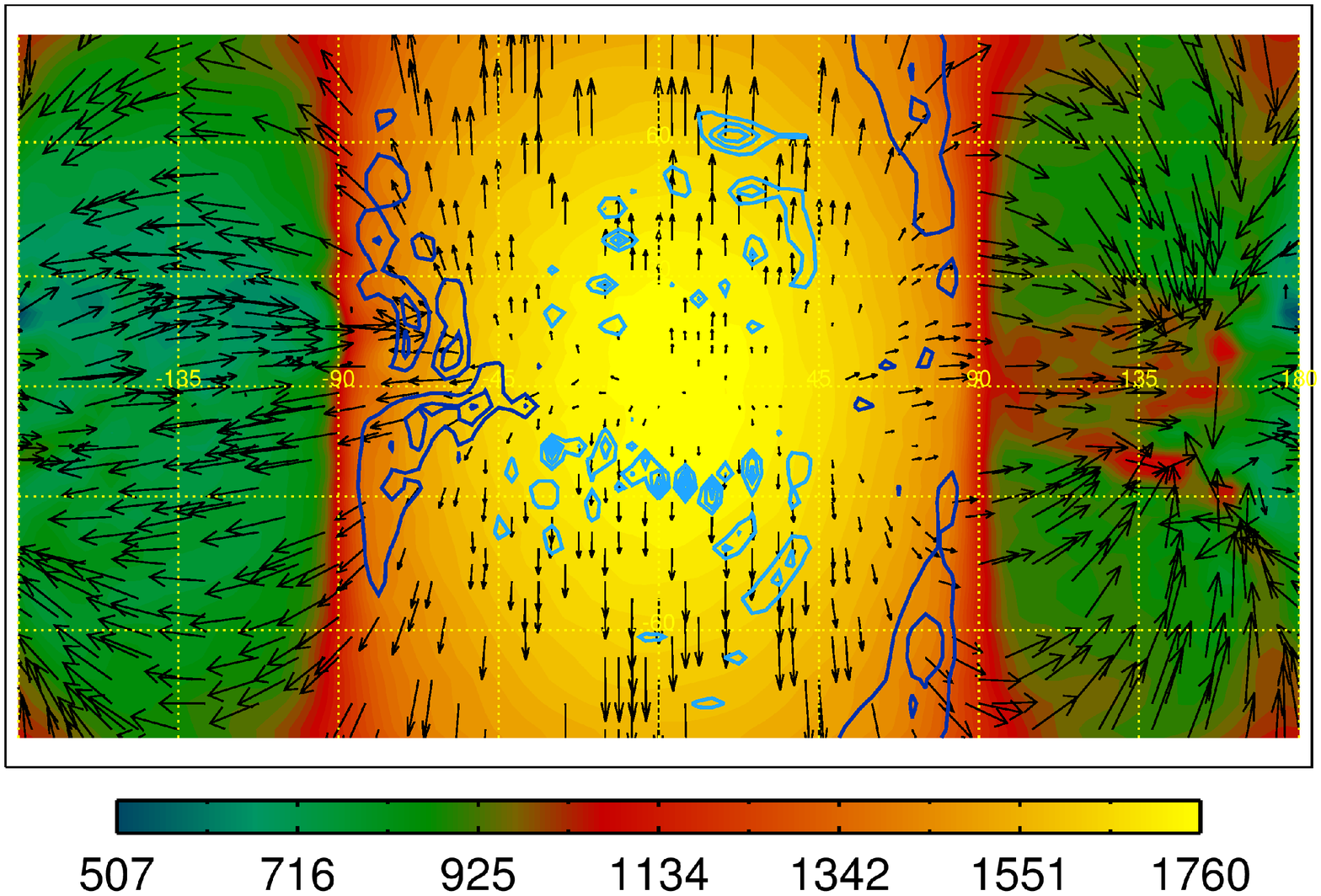}
\includegraphics[width=0.45\textwidth]{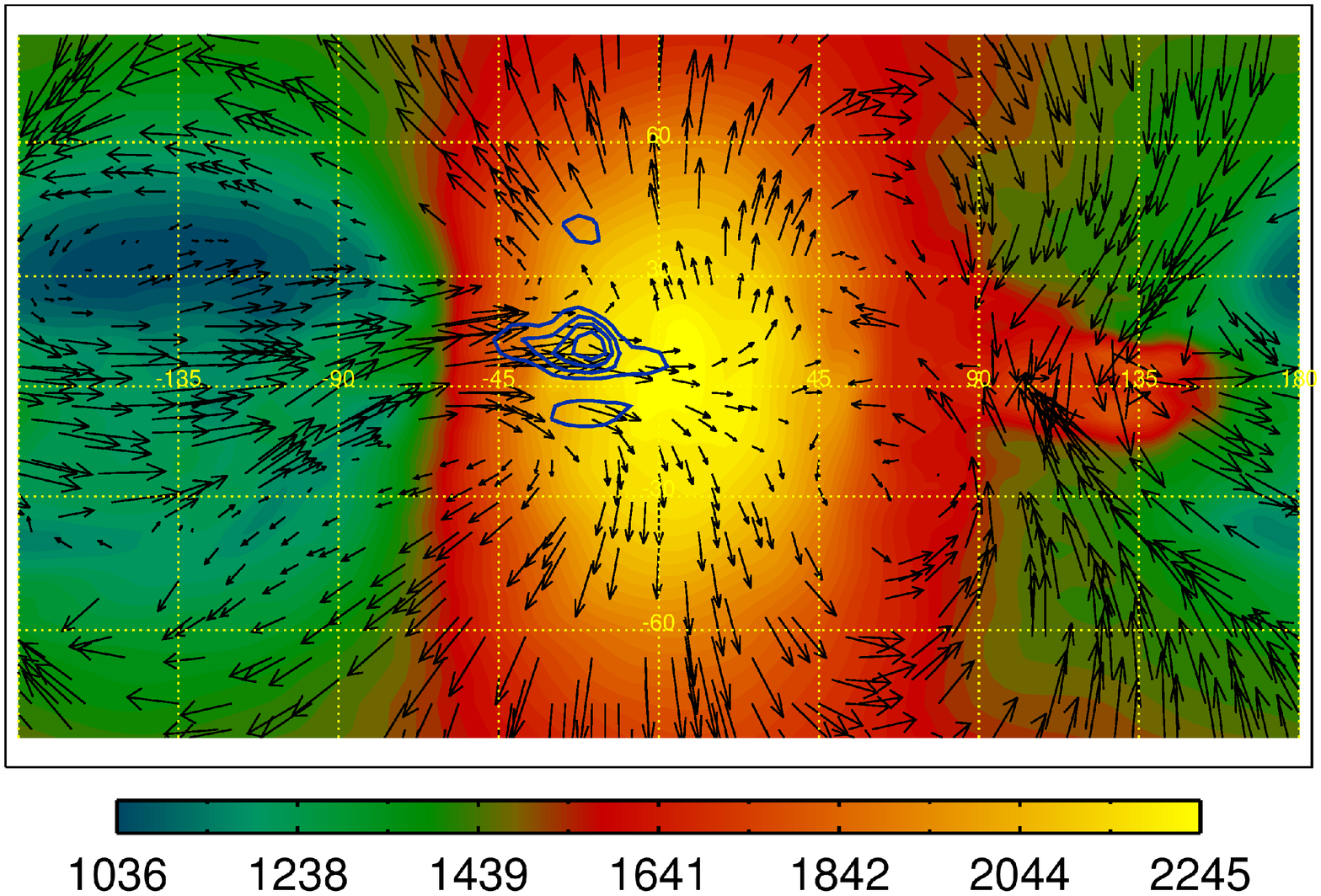}
\end{center}
\caption{Snapshots of the HD 209458b $B=30$ G model at 271 \Prot, immediately before the simulation became numerically unstable and crashed.  The dark blue contours show the ohmic heating (coincident with the magnetic drag), the light blue contours show the ohmic heating that would have occurred if we had not set a limit on its strength (\tmm$=0.005$), the color scale gives the temperature (in K) and winds are represented as arrows.  Contour levels are plotted at 20, 40, 60, and 80\% of the peak value.  Left: 3 mbar pressure level, with a peak wind speed of 8.1 \kms, a peak ohmic heating rate of 1000 \Wkg, and a peak in ``missing'' heating of 52 \Wkg.  Right: 380 mbar level, with a peak wind speed of 3.1 \kms~and a peak heating rate of 43 \Wkg.} \label{fig:r602}
\end{figure}

In Figure~\ref{fig:r602} we show layers of the HD 209458b $B=30$~G model, at 271 \Prot.  At the 3 mbar level we see the effect of artificially limiting the minimum value for \tmag~in our models.  Here the ohmic heating should be a combination of that from the zonal flow across the terminator (where winds are fast and $\tmag \geq \tmm$) and discrete regions on the day side (where zonal winds are almost nonexistent, but \tmag~is very short, $<\tmm$).  However, the lack of this heating is only a small loss; integrating the heating rates over this pressure level (see Equation~\ref{eqn:qp}), we find that it would have added only $2\times10^{18}$ W to the total ohmic heating in this pressure level, $=6\times10^{19}$ W.

The plot on the right side of Figure~\ref{fig:r602} shows a slice of the atmosphere at 380 mbar.  Here we see a significant break in hemispheric symmetry, with a perturbed equatorial jet that only exists on the night side and is strongly dragged as it tries to flow around to the day side.  In fact, the ohmic heating that results from this drag has a peak value of 43 \Wkg, which is greater than 10\% of the local rate of heating from the stellar insolation (as calculated by our radiative transfer scheme).  Below the optical photosphere, at 2 bar, there is also a region of strong ohmic heating associated with the hot feature at 135\degrees~and there the local ohmic heating is 2.8 \Wkg, or $\sim$2\% of the net radiative cooling.  Since our calculation of resistivities and magnetic timescales are updated with each timestep, the ohmic heating is strongly coupled to the dynamics and these localized regions of intense heating will fluctuate with changes in the flow, in contrast to the more steady stellar heating.  These intense local heating rates may be the cause of our numerical instability and point to the drastic influence of a $B=30$ G field on the atmospheric circulation of a planet as hot as HD~209458b.

\subsection{Comparison between HD 189733b and HD 209458b} \label{sec:hd1}

The main distinguishing feature between HD 189733b and HD 209458b, at least as applies to our models here, is the $\sim$300 K difference between their equilibrium temperatures and the resulting 2 orders of magnitude difference in their atmospheric resistivities.  This alone indicates that magnetic effects should be far less important for HD 189733b than we found them to be for HD 209458b.

\begin{figure}[ht!]
\begin{center}
\includegraphics[width=0.4\textwidth]{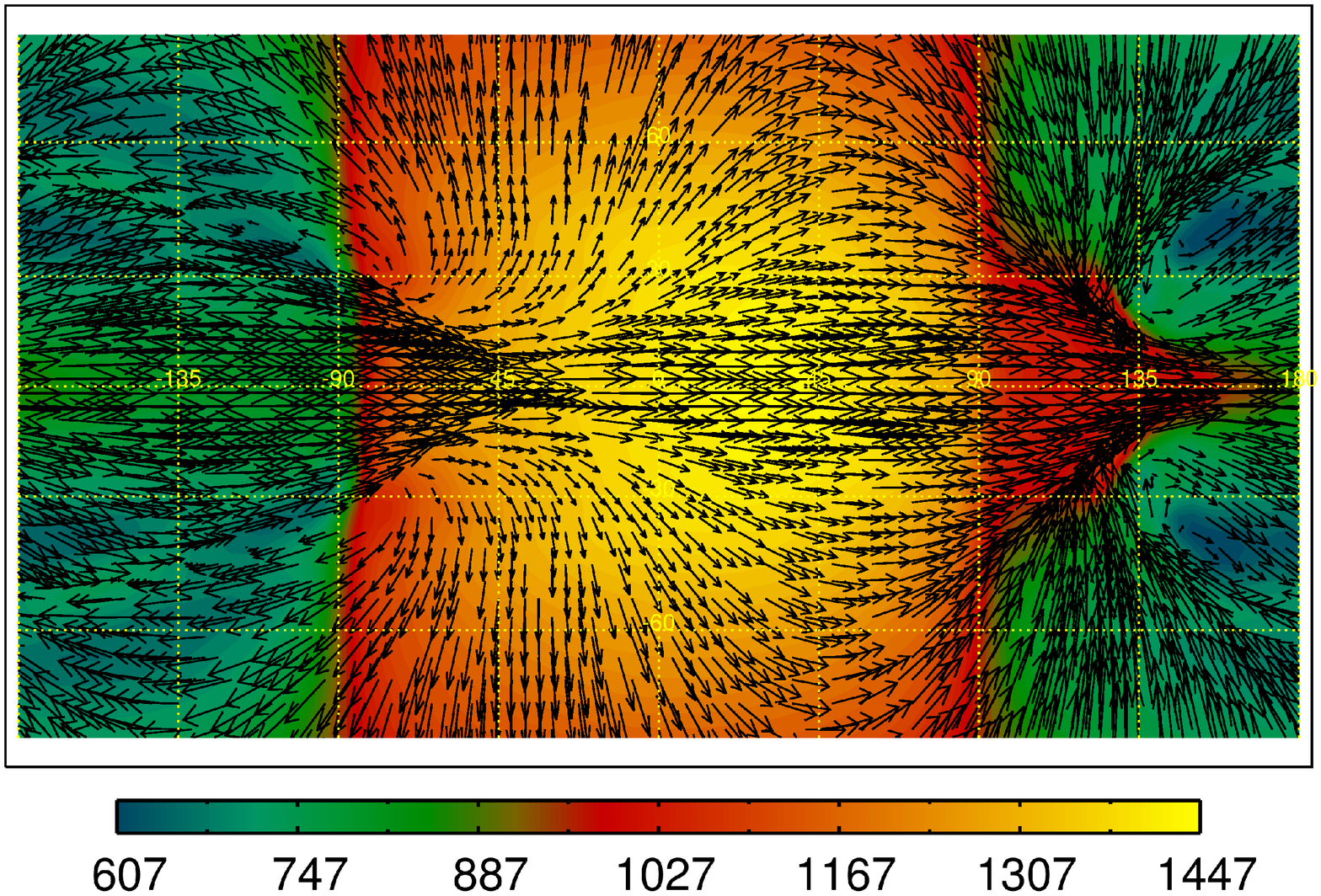}
\includegraphics[width=0.4\textwidth]{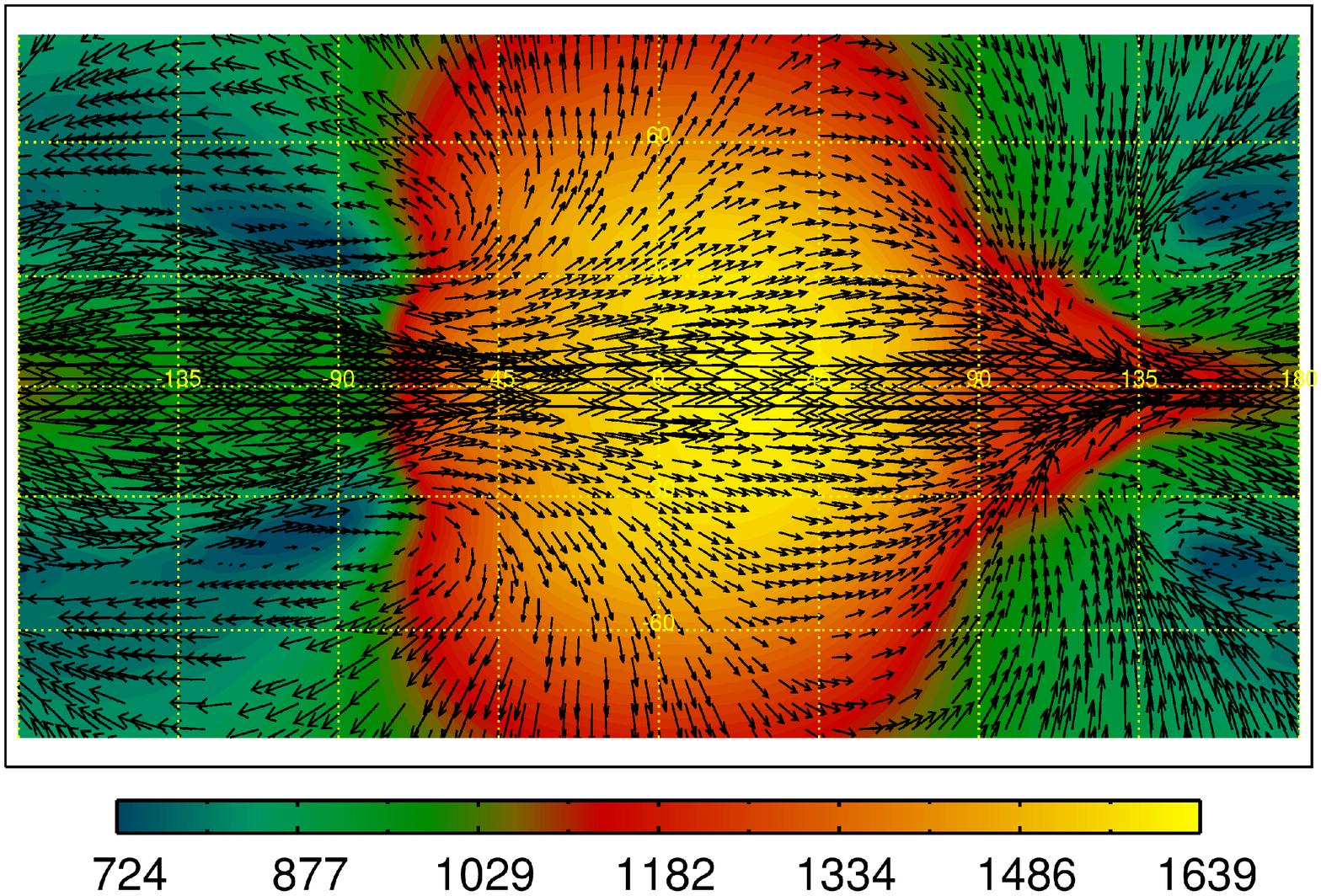}
\includegraphics[width=0.4\textwidth]{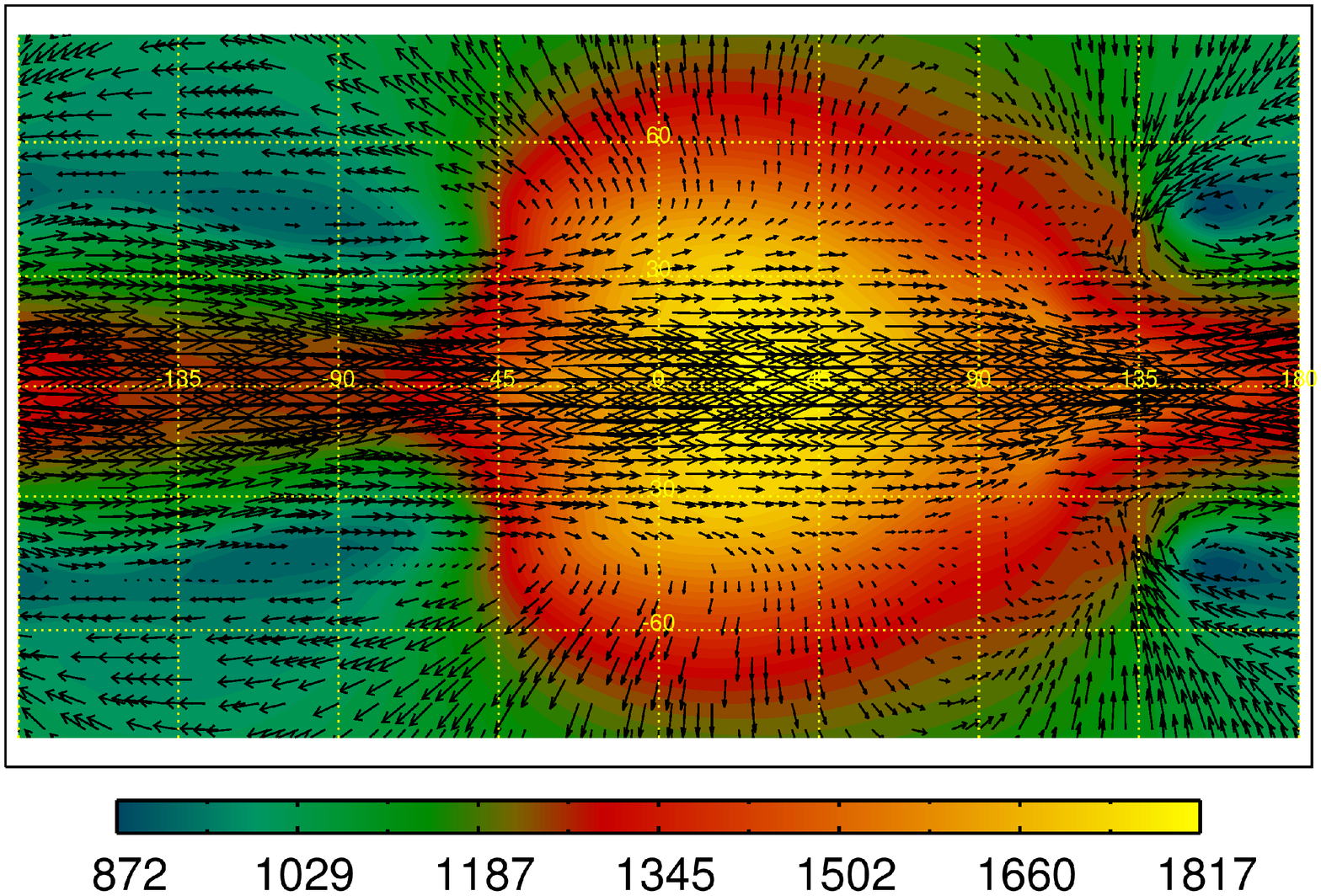}
\includegraphics[width=0.4\textwidth]{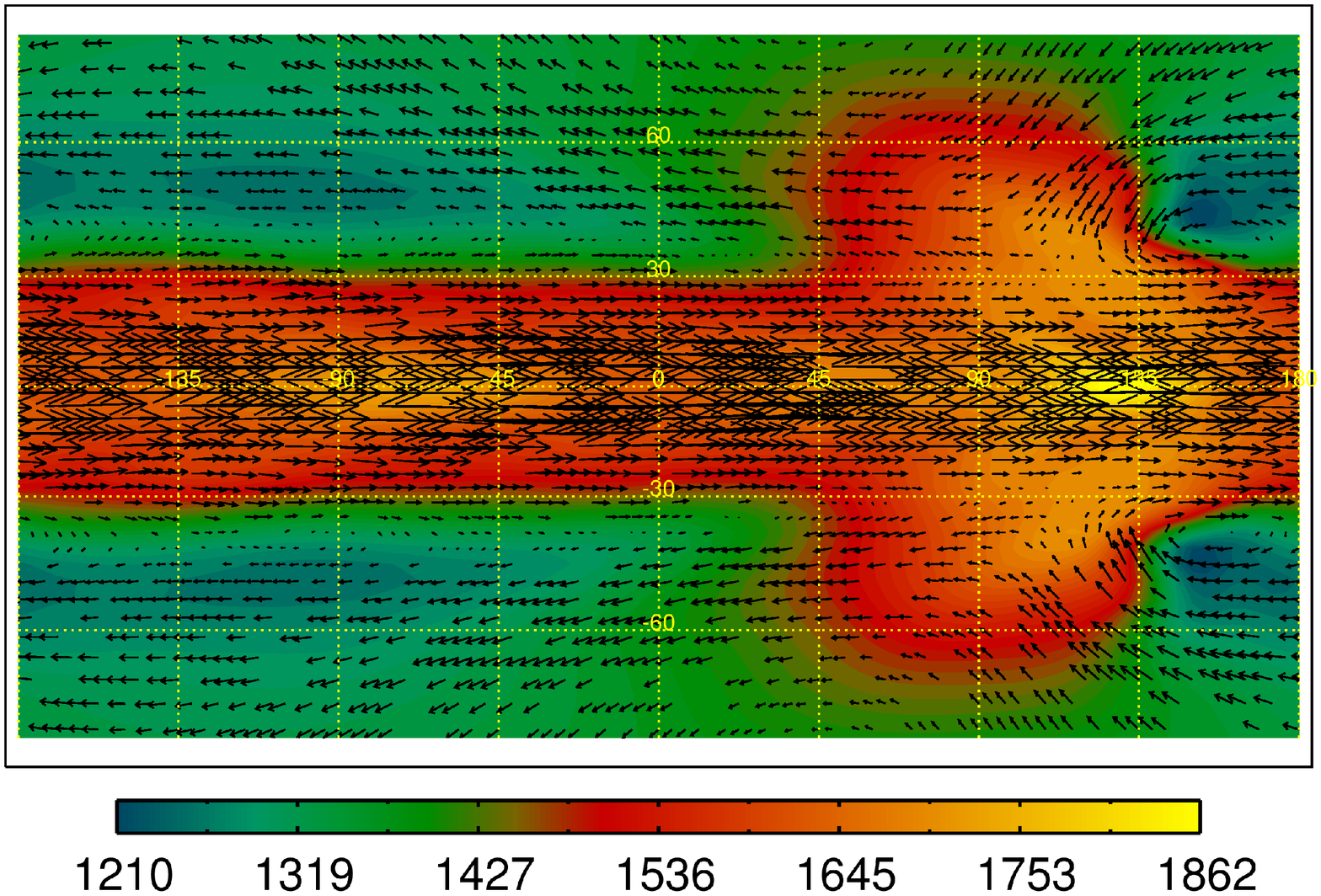}
\end{center}
\caption{Temperature and wind maps for our HD 189733b model with $B=0$ G.  The pressure levels shown are 8 mbar (top left), 180 mbar (the infrared photosphere, top right), 1 bar (bottom left), and 6 bar (bottom right), with maximum wind speeds of 7.0, 6.1, 5.2, and 4.1 \kms, respectively.} \label{fig:hd1}
\end{figure}

Since we have not previously published any non-magnetic models of HD 189733b, we first briefly present one here.  Figure~\ref{fig:hd1} shows several horizontal slices through the atmosphere, demonstrating that the circulation on HD 189733b is qualitatively similar to HD 209458b: there is a strong eastward equatorial jet that extends throughout most of the atmosphere (the zonal wind profiles are very similar, except with slower peak values for HD 189733b) and the jet becomes increasingly efficient at advecting the hottest regions away from the substellar point at deeper pressure levels.  At the photosphere this results in an eastward shift of the hot spot by $\sim$20\degrees, and by pressures of a few bars the temperatures are well homogenized around the planet.  Although the winds are slower on HD 189733b than on HD 209458b, the cooler temperatures mean that the radiative timescales are also longer, and HD 189733b has lower temperature contrasts than HD 209458b as a result.

\begin{figure}[ht!]
\begin{center}
\includegraphics[width=0.4\textwidth]{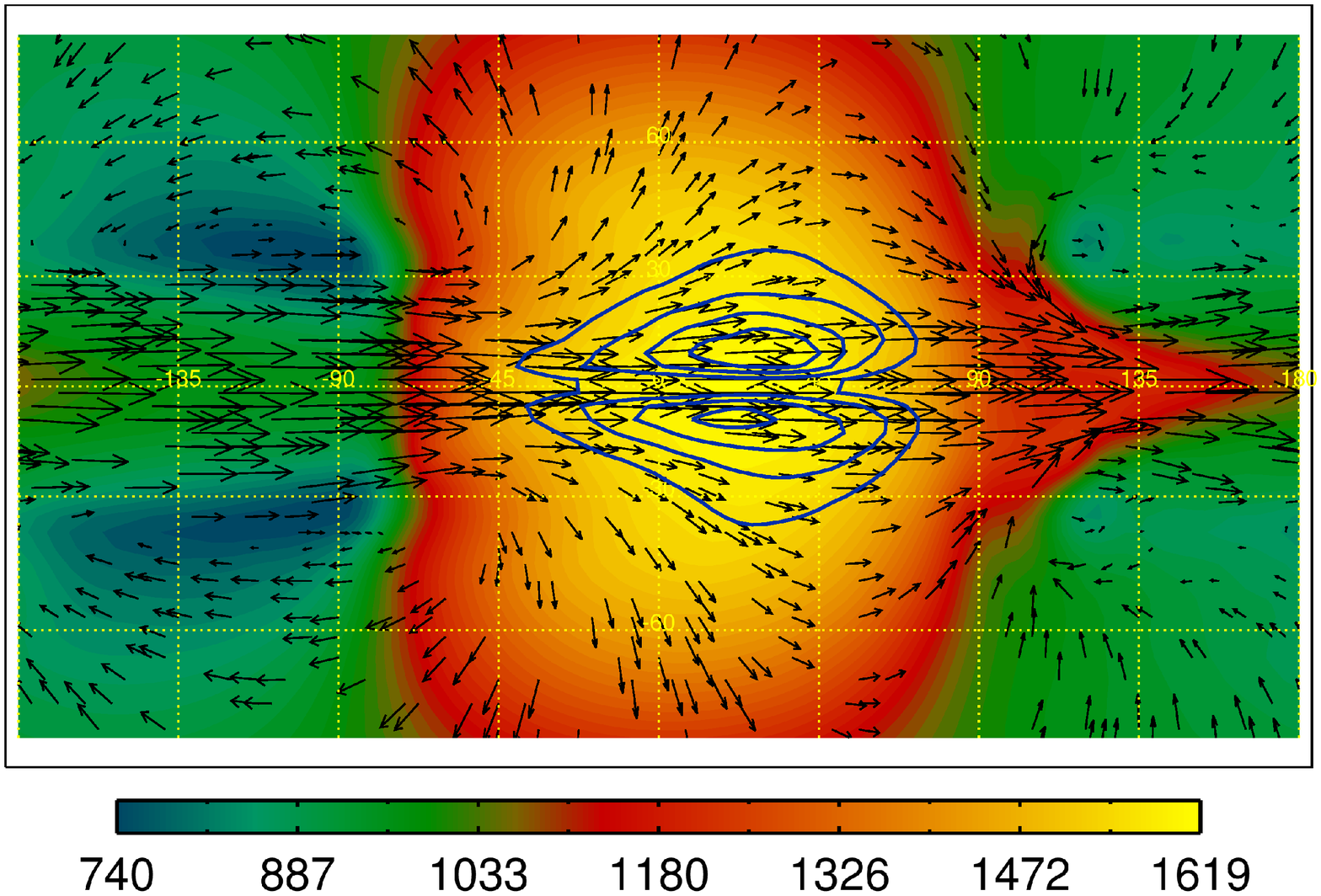}
\includegraphics[width=0.4\textwidth]{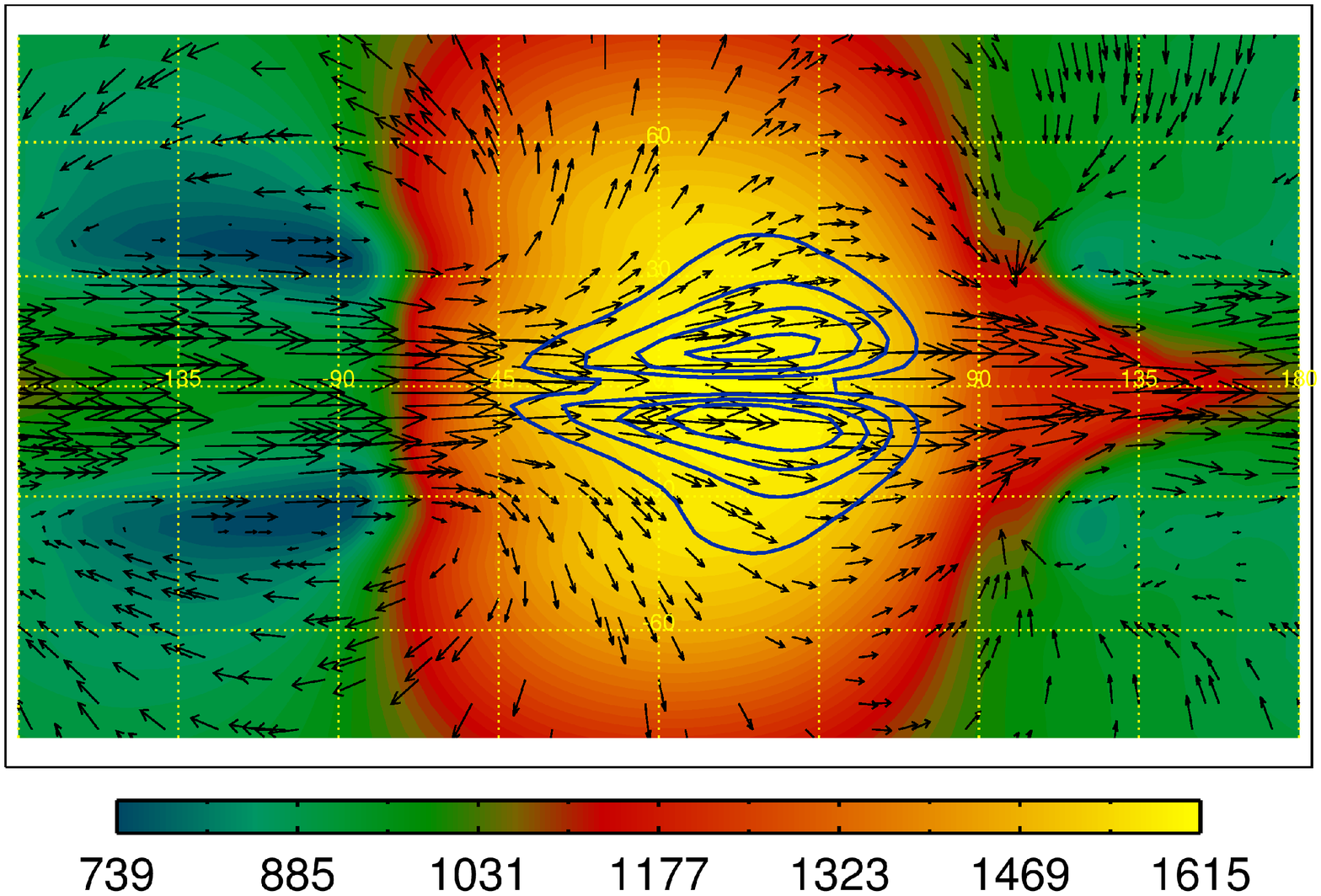}
\end{center}
\caption{Temperature, wind, and ohmic heating maps at 180 mbar for our HD 189733b model with $B=30$ G and solar metallicity (left) or 3x solar (right).  The maximum wind speeds are 5.6 and 5.3 \kms~and the peaks in ohmic heating are 5.9 and 8.4 \Wkg, respectively.}  \label{fig:hd1b30}
\end{figure}

A comparison between models of HD 189733b with $B=0$, 3, 10, and 30 G shows very minimal differences.  Wind speeds are slightly slower for higher $B$ models, but the strong equatorial jet remains and the general structure of the atmosphere is largely unchanged.  In Figure~\ref{fig:hd1b30} we plot the photospheres of models with $B=30$ G, one where we have assumed solar metallicity and one where we have used 3$\times$ solar.  Even in the case of 3$\times$ solar metallicity the atmosphere looks almost identical to the $B=0$ model.  The increase in metallicity by a factor of 3 leads to a decrease in atmospheric resistivities by $\sim$3, which is almost equivalent to an increase in the magnetic field strength by a factor of $\sqrt{3}$ (see Equation~\ref{eqn:tmag}).  Our results show that magnetic effects (as parameterized by our model) seem unable to significantly alter the atmosphere of HD 189733b, even at super-solar metallicities and for strong magnetic field strengths. 

\clearpage

\subsection{Tests of numerical resolution} \label{sec:resolution}

We might expect that the magnetic effects we have included are especially sensitive to resolution; their strength (via \tmag, Equation~\ref{eqn:tmag}) is dependent both on the latitude (sampled more frequently at higher resolution) and on the detailed temperature structure, with a small contrast in temperature resulting in an exponential difference in \tmag.  The ohmic heating and drag patterns in Figure~\ref{fig:r611} show both of these factors at work.  Often the magnetic effects are strongest near the equator, due to the higher temperatures and stronger winds commonly found there, but careful examination shows that the effects decrease sharply at the exact equator.  It is therefore beneficial to test whether the horizontal resolution of the simulation has a strong influence on the results.

However, a comparison of models at different resolutions becomes quickly complicated by the effect of hyperdissipation.  This common numerical scheme is used to reduce the build-up of energy at the smallest resolved scales by applying a high-order operator to the divergence, relative vorticity, and temperature fields.\footnote{See RM12 for a description of how we chose the hyperdissipation to use ($\tau_{\mathrm{diss}}=0.005$ \Prot~and $\nabla^8$ ) for our drag-free model of HD 209458b.}  Hyperdissipation is meant to represent subgrid processes that will continue the cascade of energy down to smaller scales, where enstrophy is eventually dissipated.  Unfortunately it is not possible to calculate the appropriate strength and order for hyperdissipation from physical principles alone; ideally it should be carefully tested for application to each particular model \citep[e.g.][]{Thrastarson2011}.  \citet{Heng2011} demonstrated that for hot Jupiter models at a given resolution, the maximum wind speeds were dependent on the choice of hyperdissipation strength, and, similarly, that maximum wind speeds vary when an identical hyperdissipation strength is applied to models at a range of resolutions.  Since magnetic effects are sensitive both to the speed of atmospheric winds and to the spatial structure of temperature and wind patterns, this makes the relationship between resolution, hyperdissipation, and magnetic effects a difficult problem to untangle.

We ran tests at lower horizontal resolution, T21 ($\sim$5.6\degrees), and a couple of limited cases at T42 ($\sim$2.8\degrees, and computationally very expensive).  Although the best choice for hyperdissipation strength should generally change with resolution, it could also depend on the magnetic effects and how they influenced the circulation pattern.  In lieu of performing a comprehensive resolution study we kept the hyperdissipation identical to that used for our T31 models.  From our resolution tests in RM12, we know that this did not result in too much over- or under-damping in our drag-free models at T21 and T42.

\begin{figure}[ht!]
\begin{center}
\includegraphics[width=0.65\textwidth]{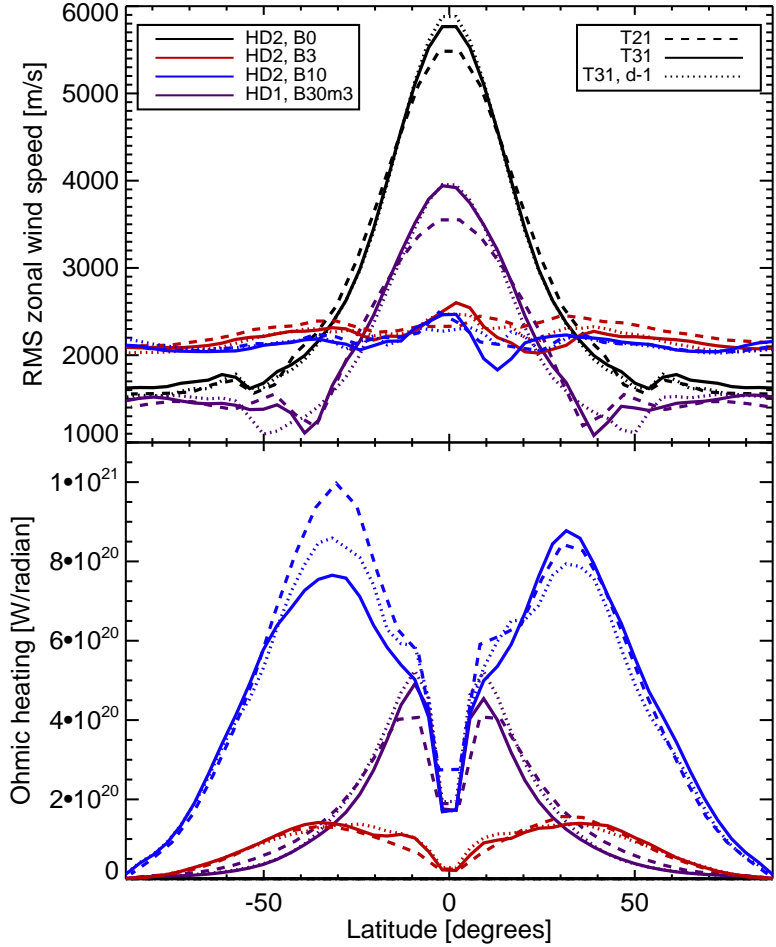}
\end{center}
\caption{Profiles of wind speeds and ohmic heating rates, as a function of latitude, for models at T21 and T31 horizontal resolutions.  The wind speeds are calculated as the root-mean-square, and the heating rates as the sum, over all longitudes and pressure levels for a given latitude.  (See Appendix A for the calculation of the heating rates.)  The profiles are for the models of HD 209458b with $B=0$ G (black), $B=3$ G (red), $B=10$ G (blue), and of HD 189733b with $B=30$ G and 3$\times$ solar metallicity (purple).  For each model we show profiles at 2000 \Prot~for T21 (dashed) and T31 (solid), with an additional profile from one rotation period earlier in the T31 runs (dotted).} \label{fig:latprofs}
\end{figure}

In general we found no significant differences between equivalent models at different resolutions; the resulting circulation patterns, temperature structures, and observable properties were fairly similar.  In Figure~\ref{fig:latprofs} we compare results from a sample of our tests, both for models in which magnetic effects were nonexistent or had a weak effect on the circulation (HD 209458b with $B=0$~G, and HD 189733b with $B=30$ G and 3$\times$ solar metallicity), and for models where the magnetic effects strongly influenced the flow (HD 209458b with $B=3$ G, and with $B=10$ G).  As expected, the models with no/weak magnetic effects have slower wind speeds in the T21 versions than at T31, due to the same hyperdissipation strength being used for both.  For the HD 189733b model this results in decreased ohmic heating near the equator, since the jet has less kinetic energy available to be dissipated through drag and converted to heat.  The HD 209458b $B=3$ G T21 model also has less near-equatorial heating than its T31 counterpart, although the difference is not as much as between the HD 189733b models.  At $B=10$ G the disparity between the T21 and T31 models of HD 209458b is comparable to the temporal variation within the T31 model.  These results indicate that as the influence of magnetic drag on the circulation increases, the influence of hyperdissipation decreases.  It also demonstrates that it is the interdependent relationship between resolution, hyperdissipation, and magnetic drag---and not each effect on its own---that influences the atmospheric circulation in these models.  While this is not an exhaustive test of numerical resolution, these results indicate that there is no sudden change in the circulation pattern at higher or lower resolution, for a range of physical properties.  Constrained by numerical cost and with the goal of testing a range of possible models, we have chosen to use T31 for the models presented in this paper.

\begin{figure}[ht!]
\begin{center}
\includegraphics[width=0.45\textwidth]{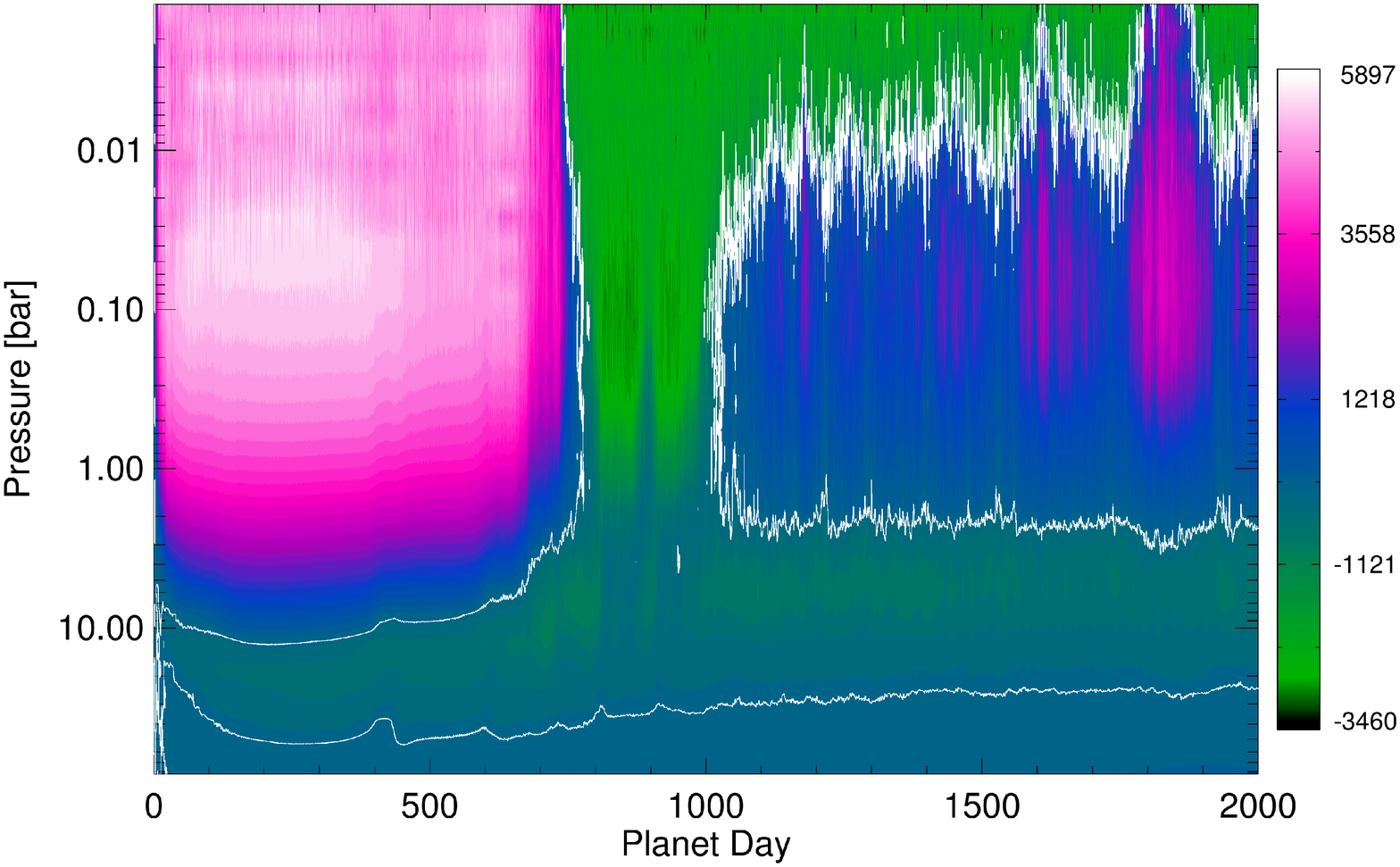} \\
\includegraphics[width=0.45\textwidth]{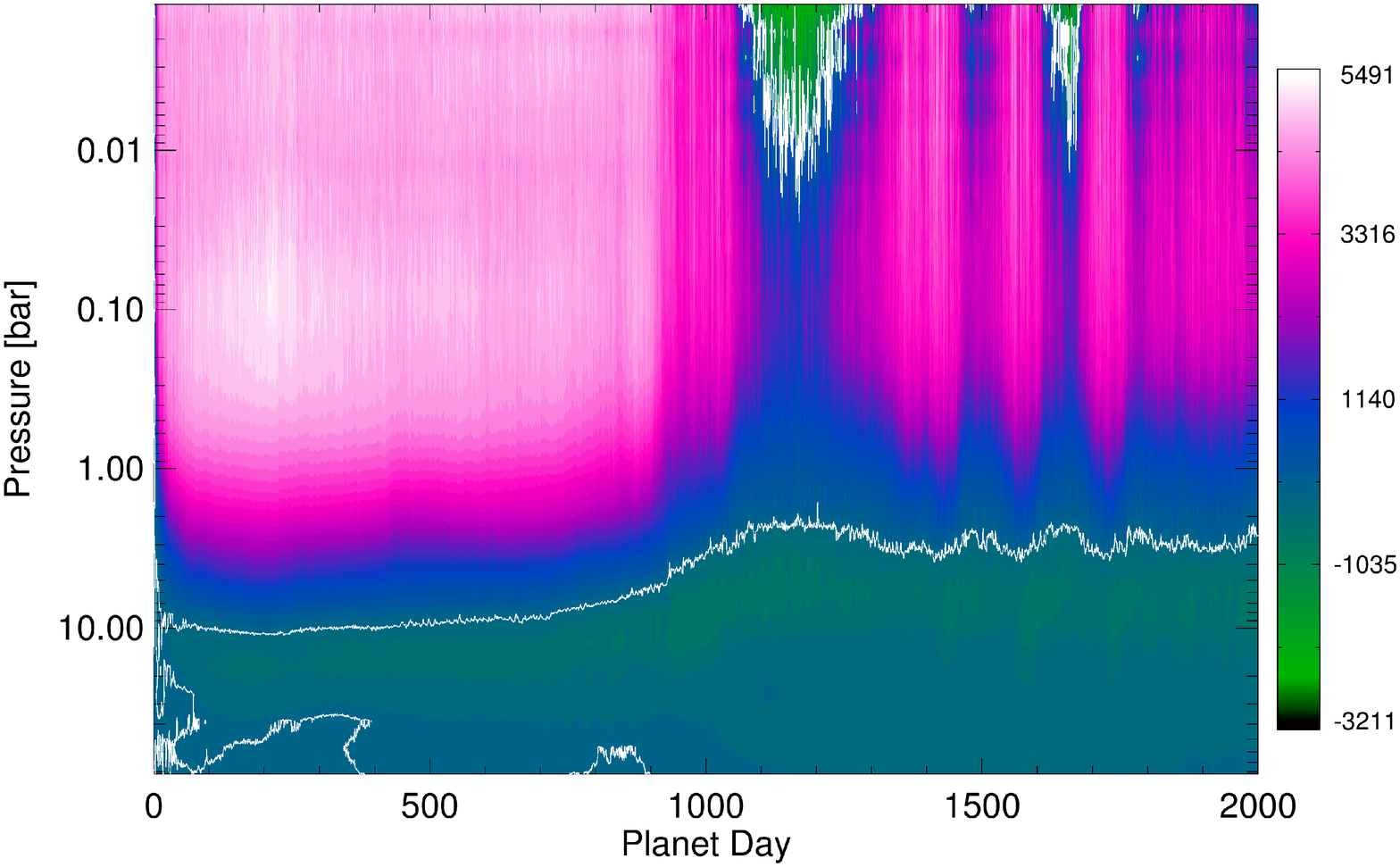} \\
\includegraphics[width=0.45\textwidth]{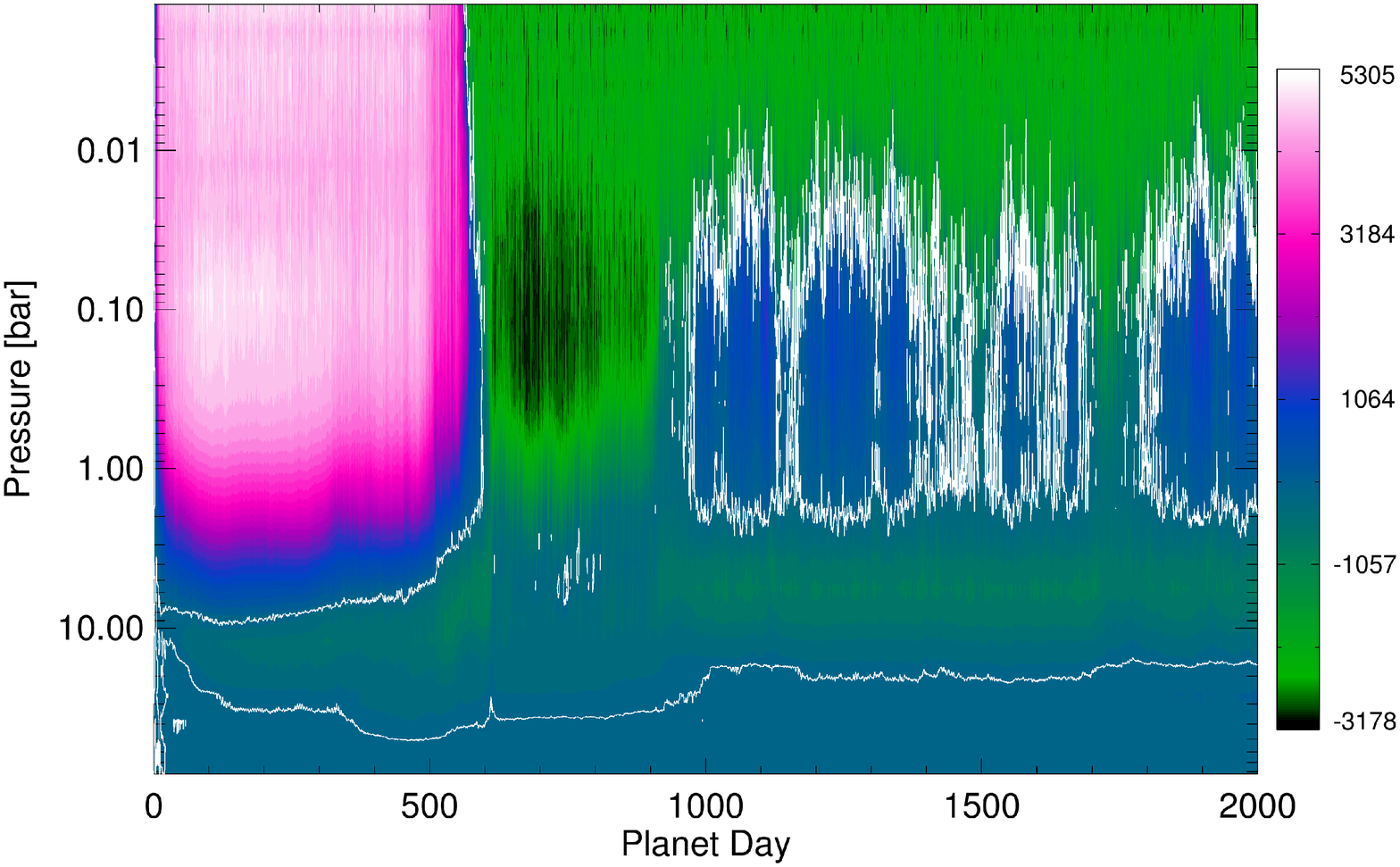}
\end{center}
\caption{Zonally averaged zonal wind speed (in \ms) at the equator, as a function of depth in the atmosphere and time in the simulation (where a ``Planet Day'' is one rotation period, \Prot).  The white line separates eastward (positive) from westward (negative) values.  The models shown are all for HD 209458b with $B=3$ G, with a horizontal resolution of T21 (top), T31 (middle), and T31 with $\Tint=100$ K (bottom).  The values at the equator are calculated as the average of $\pm1.86$\degrees~for T31 resolution and $\pm2.77$\degrees~for T21 resolution.} \label{fig:diag_uz}
\end{figure}

Although the end results were similar, we did notice an interesting peculiarity in the initial evolution of the T21 version of our HD 209458b $B=3$ G model, compared to the T31 version.  In Figure~\ref{fig:diag_uz} we plot the behavior of the zonal mean flow at the equator for models of HD 209458b with $B=3$ G, over all pressure levels, as function of time in the simulation.  The T21 and T31 versions both initially develop strong eastward equatorial flow throughout the atmosphere, which is subsequently decelerated by the magnetic drag.  In the T21 model this deceleration happens more quickly and results in a flip in the mean equatorial flow, from eastward to westward.  (This flip in the flow direction also occurs at higher latitudes, so that the dominant global flow becomes westward.)  The magnetic drag then works to oppose the westward flow and eventually the circulation returns to being primarily eastward at the equator, followed by the same irregularly periodic acceleration and deceleration of the flow seen in the T31 model.  The only other model from our set that exhibits this same flip between eastward and westward flow is our T31 model of HD 209458b with $B=3$ G and $\Tint=100$ K (also shown in Figure~\ref{fig:diag_uz}).  We emphasize that the final circulation states of these models are all similar, but we present this peculiarity in order to demonstrate that the presence of magnetic effects can influence not just the state of atmospheric circulation, but also its development during a simulation.  The inclusion of magnetic effects expands the dynamical possibilities for the atmospheric flow and introduces greater variability in its development.  
Given that we have made simplifying assumptions in order to begin to explore the influence of magnetic effects, the diversity of behavior may well be much greater than we have seen here.
The potential richness of expanded dynamical regimes will be explored in future work.

\clearpage

\subsection{Effects on observable properties} \label{sec:obs}

Most of the observable properties of an exoplanet can contain signatures of its atmospheric circulation.  As the planet transits the star, the detailed shape of the transit curve will depend on the geometric shape of the planet, which can be altered by the atmospheric circulation \citep{Barnes2009,DobbsDixon2012}.  The depth of the transit as a function of wavelength will depend on the composition and temperature profiles along the terminator, a region particularly vulnerable to the influence of winds blowing from day to night \citep{Burrows2010,Fortney2010,DobbsDixon2012}.  With future observational facilities it may be possible to directly measure the speed of these winds, as they produce a Doppler shift in spectra taken during transit \citep{Snellen2010,Kempton2012,Showman2012}.  The atmospheric circulation can even have an indirect effect on the size of the planet, via heating by ohmic dissipation, as we discuss in the next section.

\begin{figure}[ht!]
\begin{center}
\includegraphics[width=0.32\textwidth]{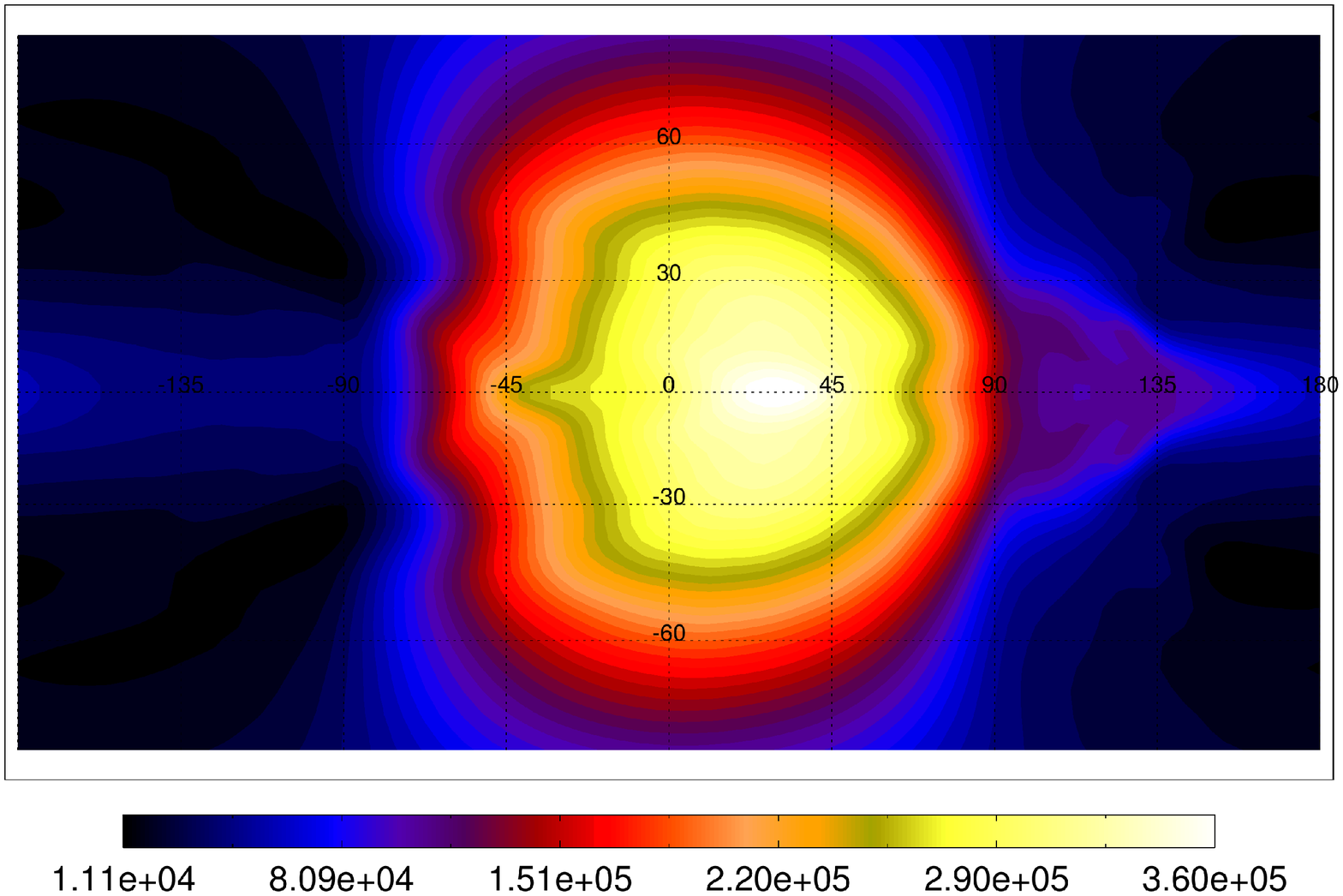}
\includegraphics[width=0.32\textwidth]{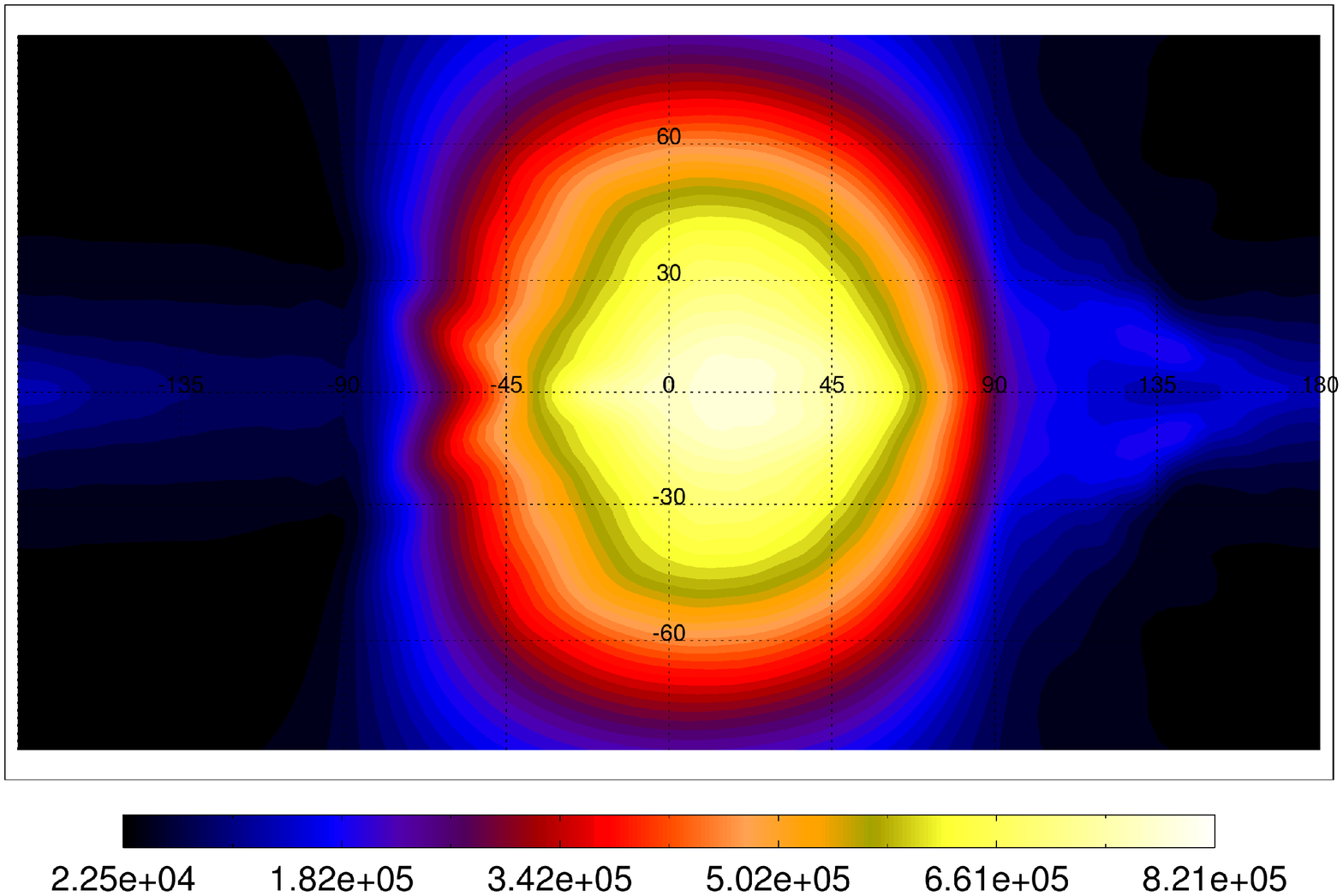} \\ 
\includegraphics[width=0.32\textwidth]{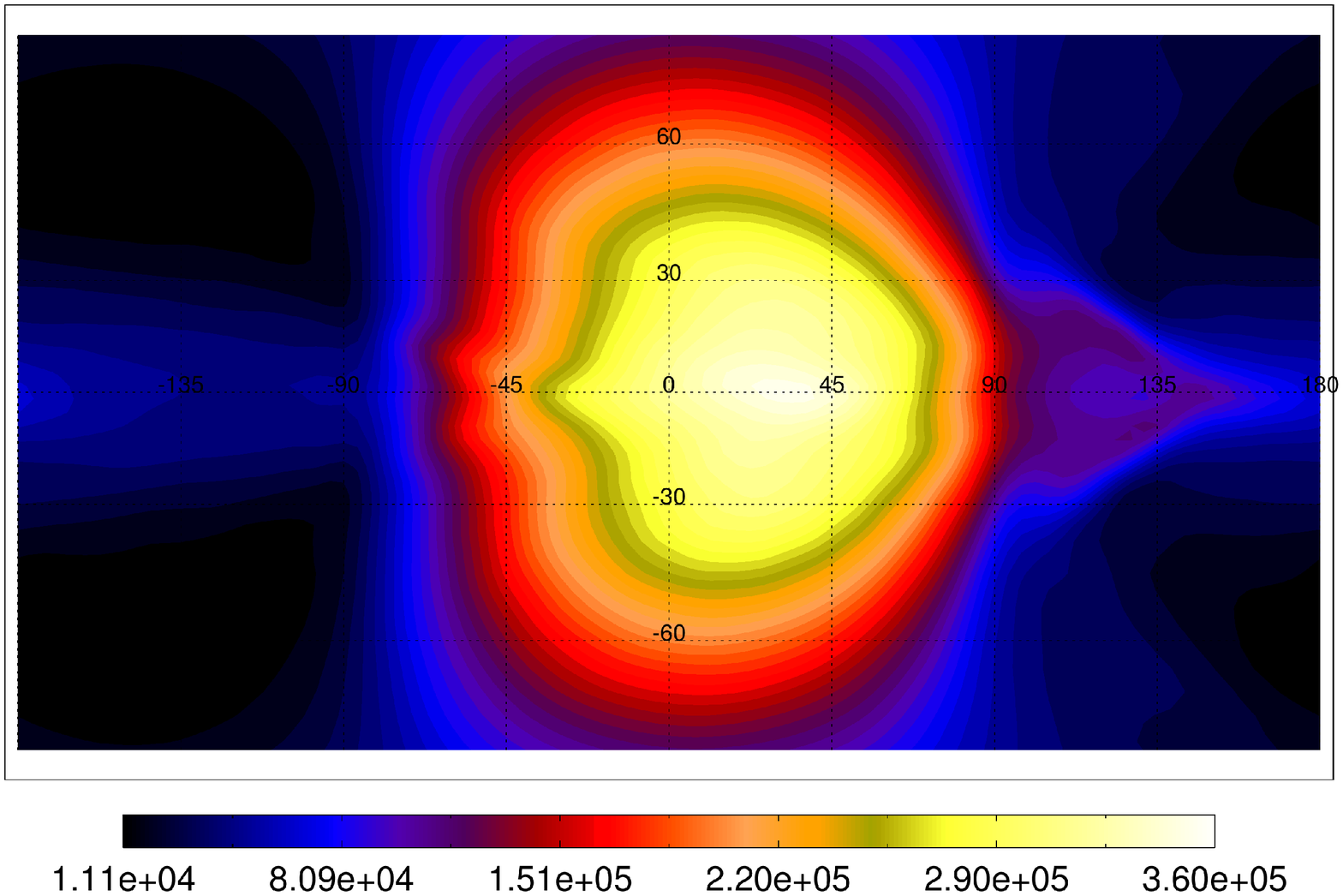}
\includegraphics[width=0.32\textwidth]{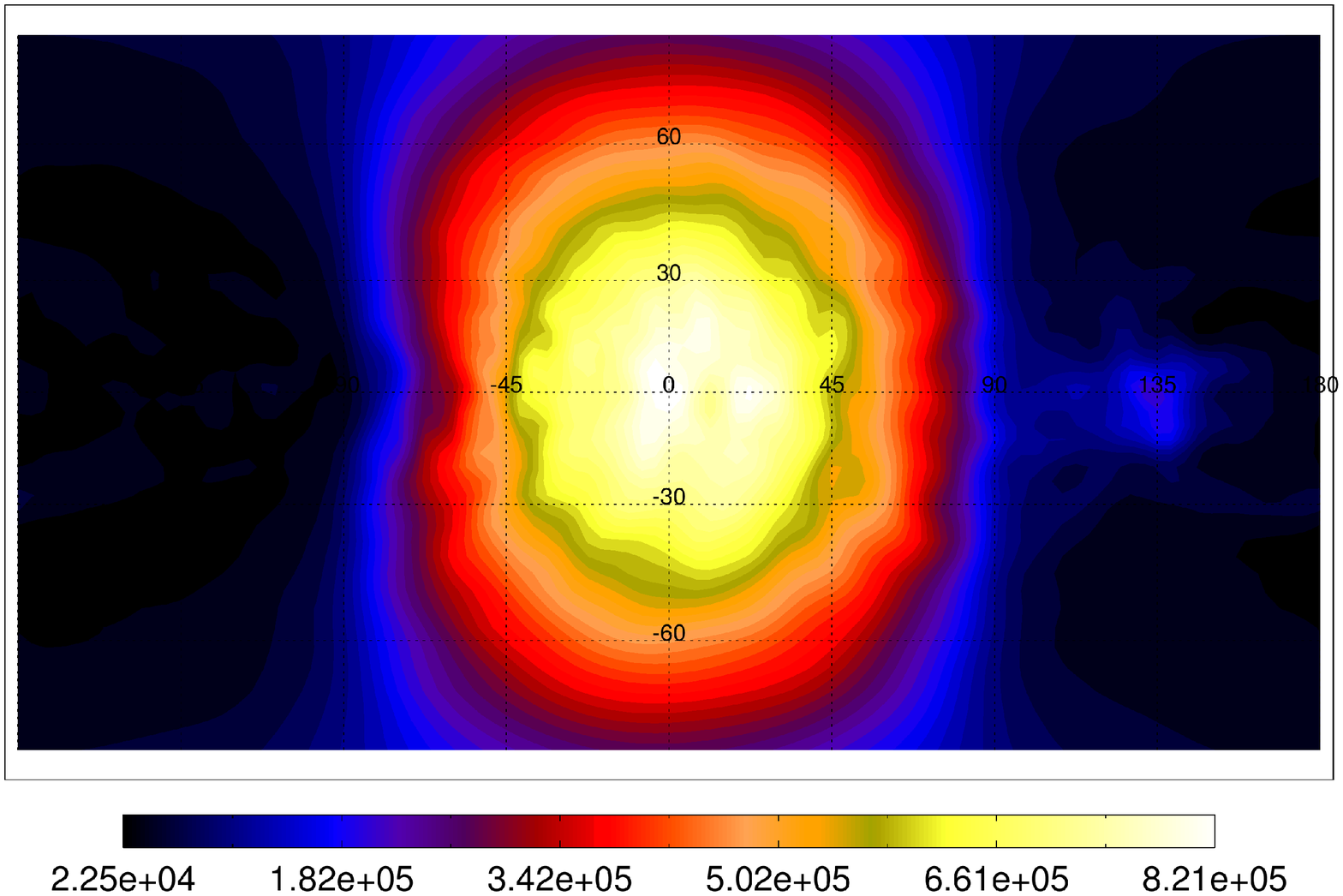} \\
\includegraphics[width=0.32\textwidth]{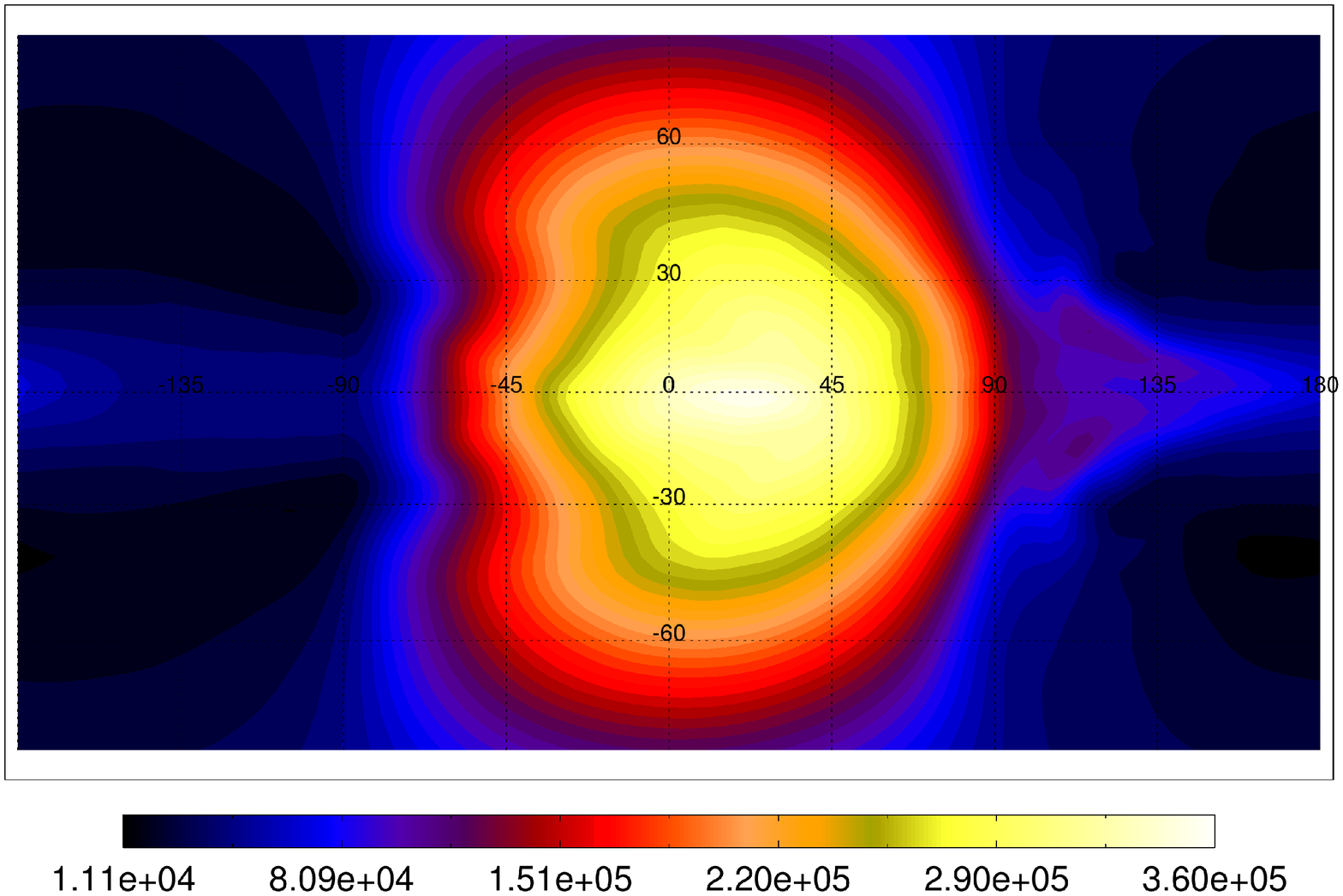} 
\includegraphics[width=0.32\textwidth]{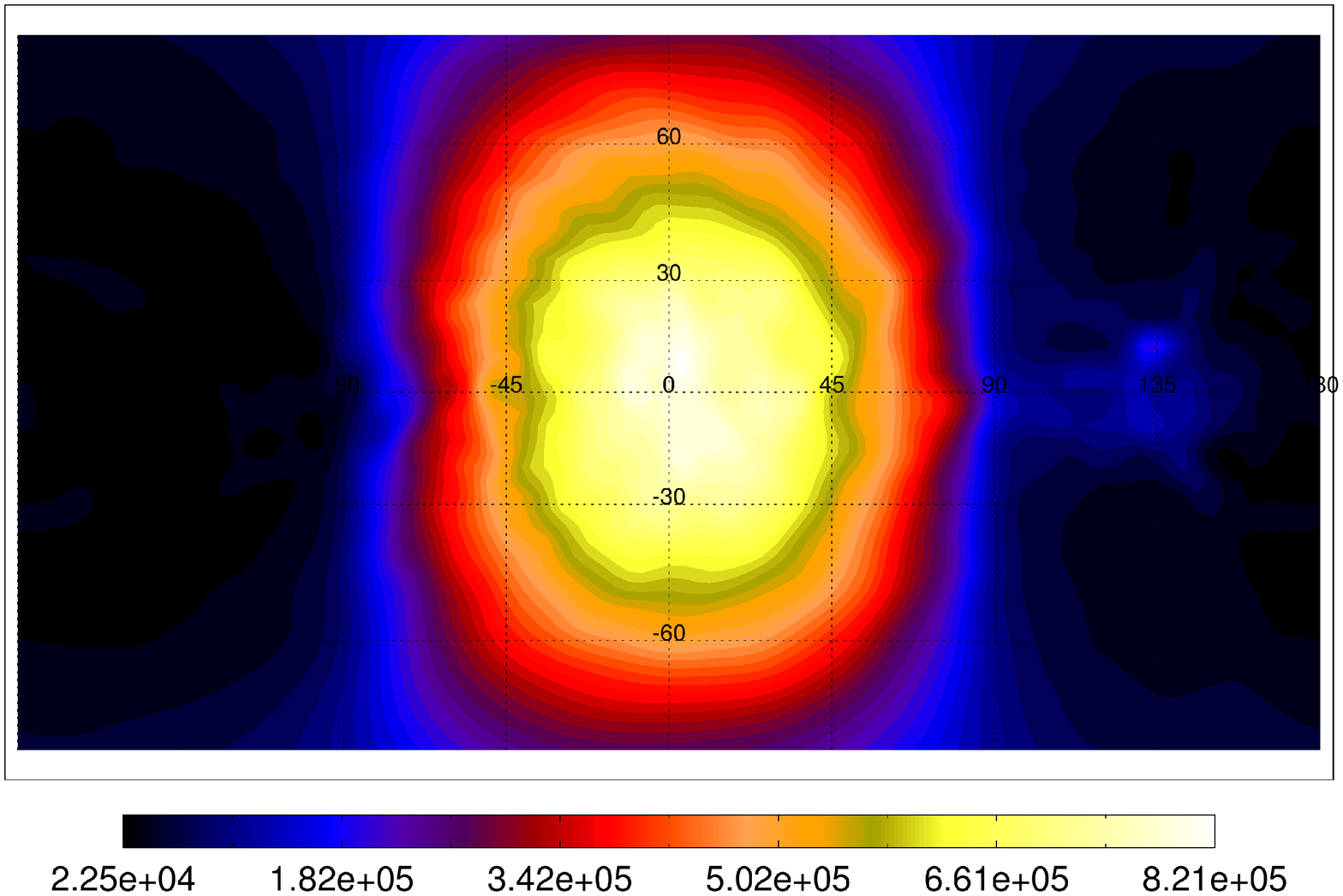} \\
\includegraphics[width=0.32\textwidth]{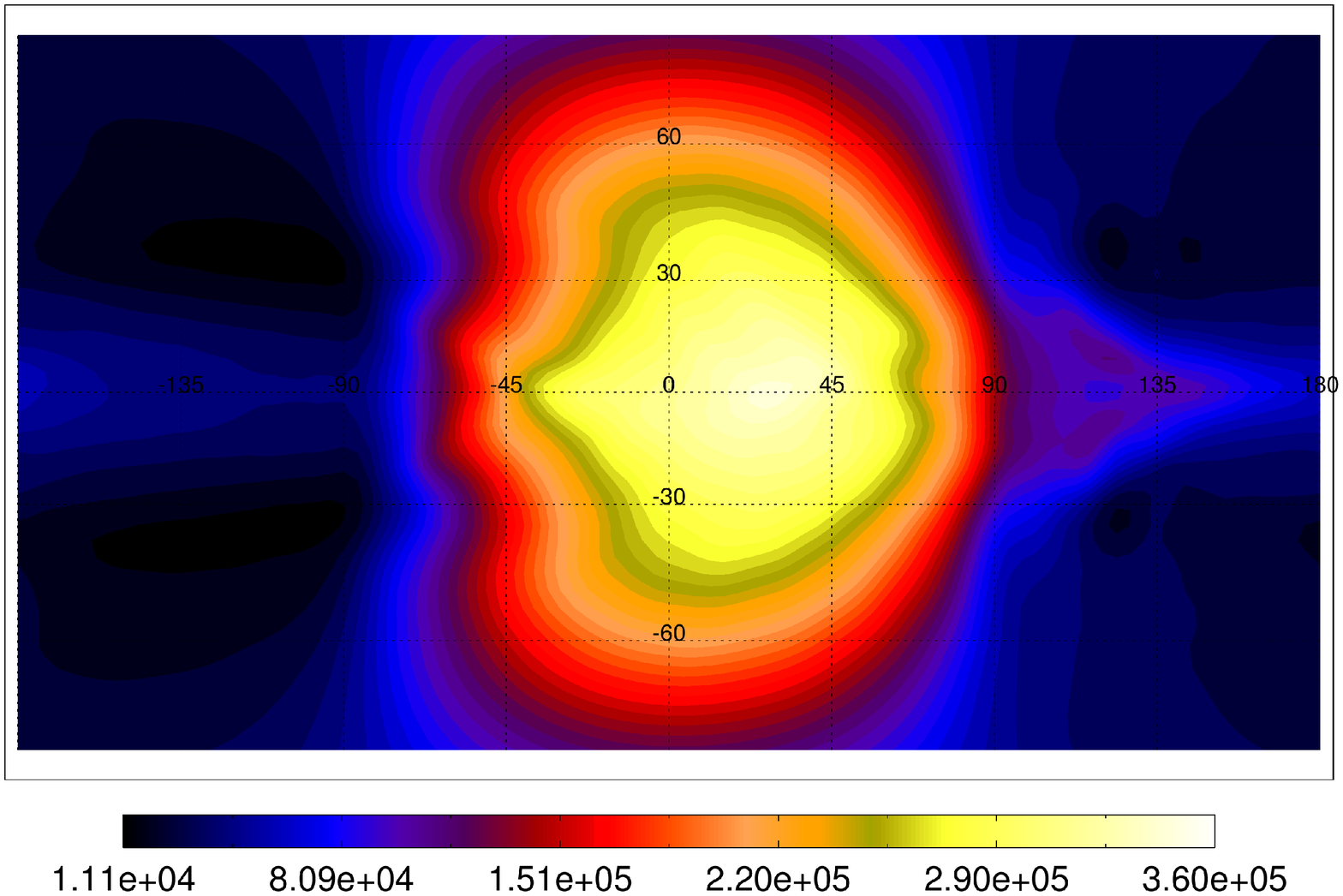}
\includegraphics[width=0.32\textwidth]{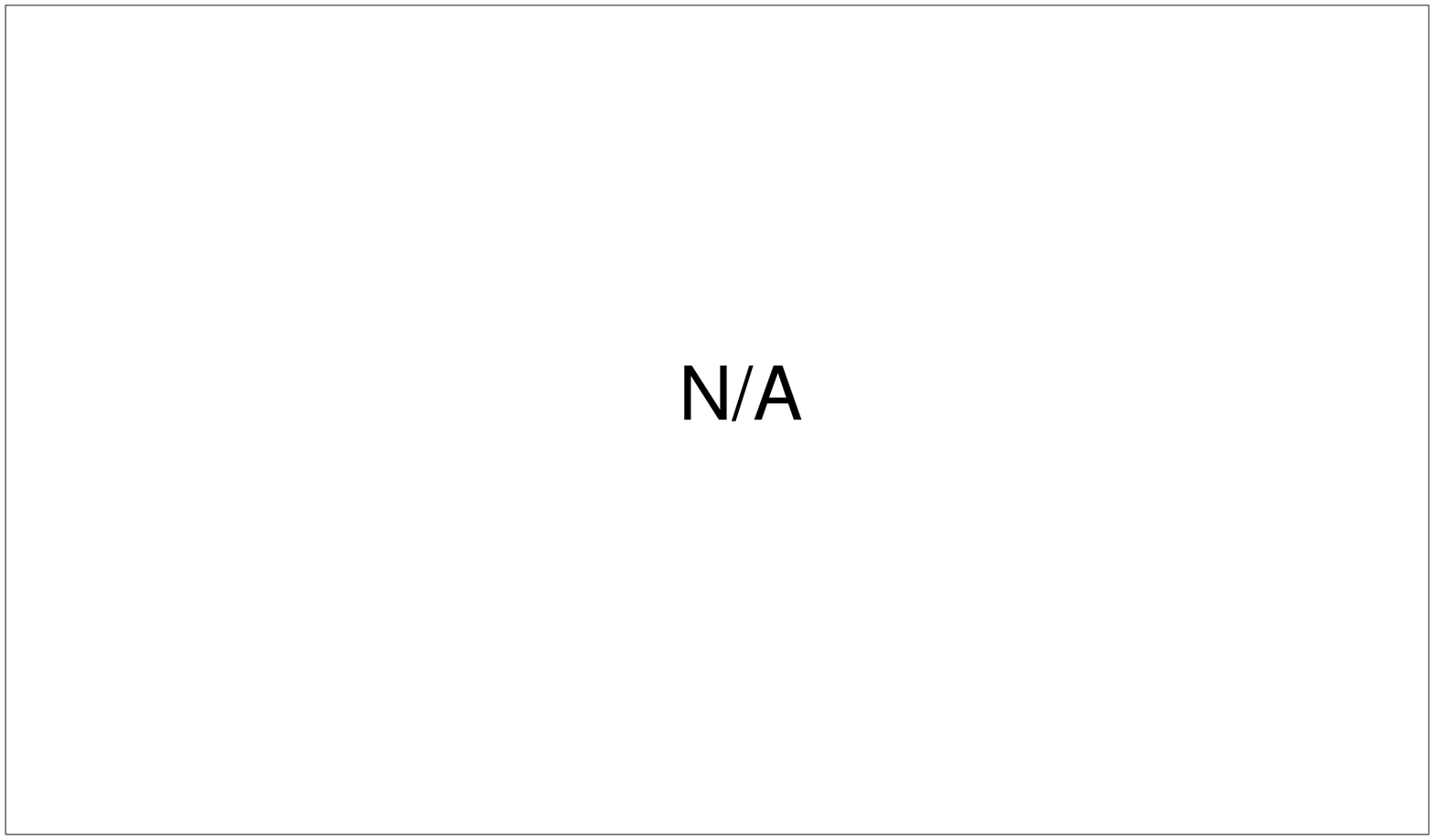}
\end{center}
\caption{Cylindrical maps of the infrared flux (in \Wms) emitted from the top boundary of the models for HD 189733b (left column) and HD 209458b (right column) with $B=0$ G (top row), $B=3$ G (second row), $B=10$ G (third row), and $B=30$ G (bottom row).  The substellar point is at the center of each plot.}  \label{fig:drag_folr}
\end{figure}

\begin{figure}[ht!]
\begin{center}
\includegraphics[width=0.85\textwidth]{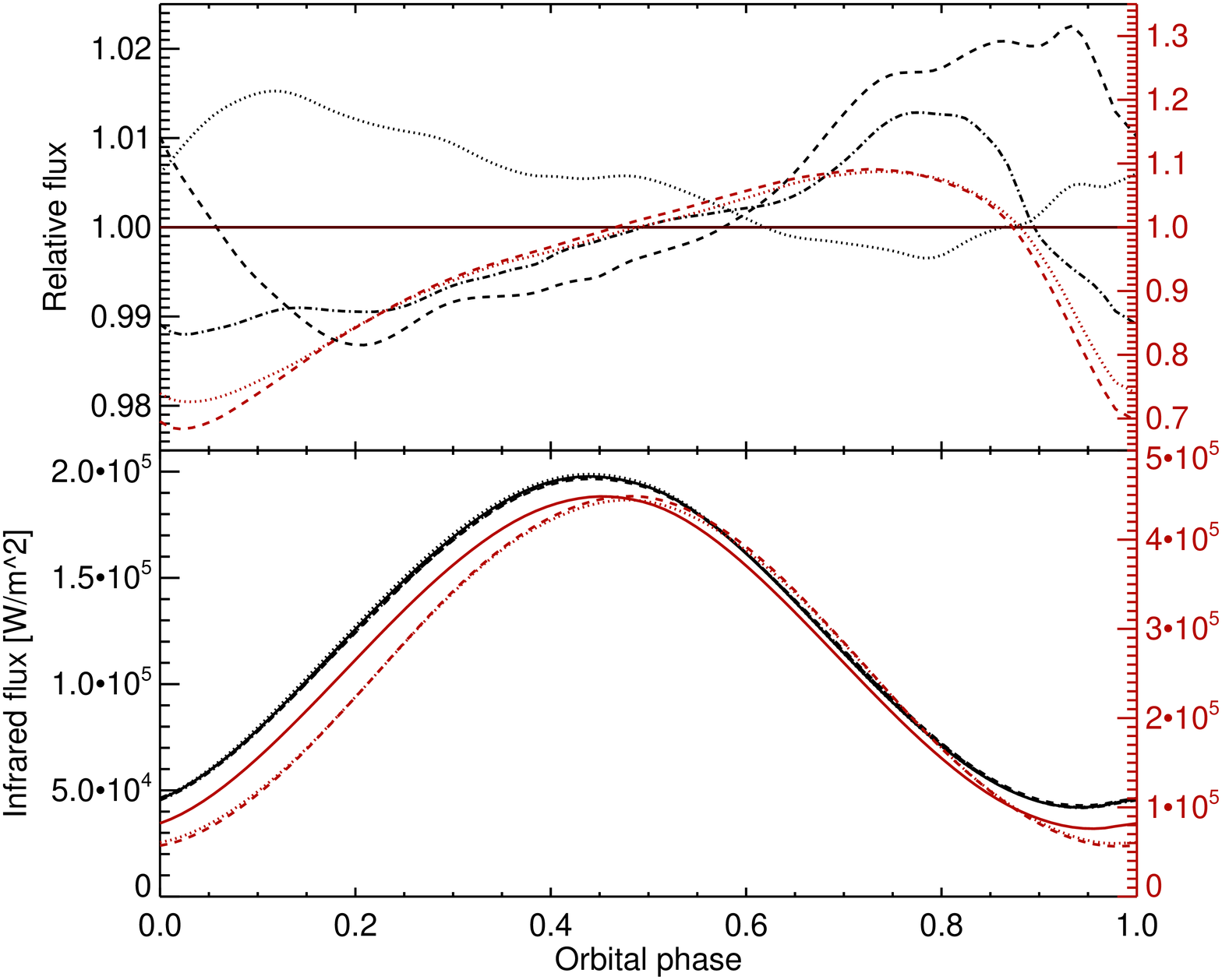}
\end{center}
\caption{Emitted infrared phase curves for models of HD 189733b (black) and HD 209458b (red) with $B=0$ G (solid), $B=3$ G (dotted), $B=10$ G (dashed), and $B=30$ G (dash-dotted).  The bottom panel gives the emitted flux, averaged over the hemisphere facing the observer, while the top panel gives the ratio between each magnetic model and the $B=0$ G model.  The planet transits the star at a phase of 0 and passes behind the star at a phase of 0.5; the dip in light during secondary eclipse (not shown) provides a measure of the emission from the day side of the planet.} \label{fig:pc}
\end{figure}

In addition to observing the shadow of the planet (and its atmosphere) during transit, we can also measure the light emitted by the planet, which for hot Jupiters typically peaks in the infrared.  This is a property of the planet that is self consistently predicted by our numerical code.  In Figure~\ref{fig:drag_folr} we plot maps of the infrared flux emitted from the top of the atmosphere,\footnote{Note that in our modeling scheme all of the thermal emission of the planet is captured in a single band, so that we have no wavelength information.} for models of HD 189733b and HD 209458b with $B$ from 0 to 30 G.  It is immediately apparent from these maps that magnetic effects significantly change the observable properties of HD 209458b, but not necessarily those of HD 189733b.  We integrate these maps over the planet disk, for viewing orientations along the equator, to calculate the change in infrared emission that would be measured by a distant observer as the planet orbits its star, assuming that the planet's orbital and rotation periods are the same (as we have in our model set-up).  These orbital phase curves are plotted in Figure~\ref{fig:pc}, where we also plot the phase curve for each magnetic model divided by the curve for the $B=0$ G model of that planet, in order to see small differences between similar curves.  

As expected from the maps in Figure~\ref{fig:drag_folr}, we see a significant change between the $B=0$ G  and $B>0$ G models of HD 209458b.  The $B=0$ model has an effective longitude of peak emission that is shifted away from the substellar point (by $\theta_{\mathrm{Fmax}}=12$\degrees, as reported in Table~\ref{tab:obs}), while the $B>0$ models all have their peak emission well aligned with the substellar point ($\theta_{\mathrm{Fmax}} = 0-3$\degrees).  There is a lower ratio between the minimum and maximum flux for the $B>0$ models than the $B=0$ one ($F_{\mathrm{min}}/F_{\mathrm{max}}=12-13$\%, compared to 17\%) and the differences between these models are greater than the temporal variation in emission within any of the models, which is at a level of less than 1\%.  On the other hand, there is practically no difference between the phase curves for our various models of HD 189733b.  They all have almost the same effective longitude of peak emission ($\theta_{\mathrm{Fmax}}=17$\degrees) and flux ratio between minimum and maximum emission ($F_{\mathrm{min}}/F_{\mathrm{max}}=21$\%).  The relative differences between the phase curves of these models is comparable to the low level of temporal variation in emission within each model ($\lesssim$2\%).

\clearpage
\begin{deluxetable}{lccc}
\tablewidth{0pt}
\tablecaption{Observable properties of each model}
\tablehead{
\colhead{Model} & \colhead{$F_\mathrm{min}/F_\mathrm{max}$} 
	& \colhead{$\theta_{\mathrm{Fmax}}$} 	& \colhead{$\varepsilon$}
}
\startdata
HD 209458b, with:			&	&	&	\\
\ \ \ $B=0$ G				& 17\%	& 12\degrees	& 0	\\
\ \ \ $B=3$ G, $\Tint=100$ K	& 13\%	& 0\degrees	& 0.7\% 	\\
\ \ \ $B=3$ G				& 13\%	& 3\degrees	& 0.7\%	\\
\ \ \ $B=10$ G				& 12\%	& 2\degrees	& 4\%	\\
\ \ \ $B=30$ G\tablenotemark{a}& n/a	& n/a 		& 10\%	\\
\hline
HD 189733b, with:			&	&	& 	\\
\ \ \ $B=0$ G 				& 21\%	& 17\degrees	& 0	\\
\ \ \ $B=3$ G 				& 21\%	& 17\degrees	& 0.03\%	\\
\ \ \ $B=10$ G				& 21\%	& 17\degrees	& 0.3\%	\\
\ \ \ $B=30$ G				& 21\%	& 16\degrees	& 2\%	\\
\ \ \ $B=30$, metallicity$\times 3$ & 20\%	& 16\degrees	& 3\%	\\
\enddata
\label{tab:obs}
\tablenotetext{a}{This is from the last snapshot before the model crashed at 271 \Prot~and so is not a robust result, but we include the ohmic heating for comparison with other models.}
\tablecomments{The observable properties of each model are: the flux ratio between minimum and maximum emission ($F_\mathrm{min}/F_\mathrm{max}$), the longitude of maximum emission ($\theta_{\mathrm{Fmax}}$), and the total ohmic heating as a fraction of the stellar input ($\varepsilon$).  In the calculation of $\varepsilon$, the total luminosity of incident stellar irradiation is $3.3 \times 10^{22}$ W for HD 209458b and $1.5\times 10^{22}$ W for HD 189733b.  All quantities are calculated from a snapshot of the atmosphere at 2000 \Prot.}
\end{deluxetable}

Even though HD 209458b is one of the brightest hot Jupiters known, it has no high quality phase curve measurements published, and the lack of a wavelength dependent radiative transfer routine in our numerical code precludes any detailed comparison with available multi-wavelength secondary eclipse measurements.  If taken at face value, the observations of HD 209458b by \citet{Cowan2007} would indicate an infrared flux ratio ($F_{\mathrm{min}}/F_{\mathrm{max}}$) of less than $0.15-1.5$\%; however, this is calculated from several discrete observations, a much less reliable way to measure a phase curve than continuous observation \citep[compare][]{Harrington2006,Crossfield2010}.  \citet{Crossfield2012} reported that an existing continuous phase curve observation of HD 209458b at 24\micron~is too plagued by instrumental effects to provide a reliable measurement.

Fortunately, there are published phase curve results for HD 189733b at a fairly high level of precision.  \citet{Knutson2012} report multi-wavelength continuous phase curve observations of HD 189733b, from 3.6 to 24\micron~\citep[including previous results from][]{Knutson2007,Knutson2009}.  Since hot Jupiters are not blackbodies, but in fact have strong molecular bands in the infrared, there are significant differences between the phase curves at each wavelength.  The ratio between minimum and maximum flux ranges from $F_{\mathrm{min}}/F_{\mathrm{max}}=15-74$\% and the range for the offset of the hot spot from the substellar point is $\theta_{\mathrm{Fmax}}=20-37$\degrees.  The work by \citet{Agol2010} finds values that fall within this range and also constrains the temporal variation in 8\micron~day side emission to be less than 2.7\% (at 68\% confidence).  

Although our model results can only roughly agree with measurements, due our lack of multi-wavelength radiative transfer, it is helpful for us to compare the \emph{precision} of these measurements to amount by which we predict that magnetic effects should be able to influence observable properties.  In \citet{Knutson2012} they are able to achieve precisions as good as $\pm5$\% for $F_{\mathrm{min}}/F_{\mathrm{max}}$ and $\pm3$\degrees~for $\theta_{\mathrm{Fmax}}$, for observations of HD 189733b.  We do not expect magnetic effects to strongly influence the circulation on this planet.  However, the signature of magnetic effects in the atmosphere of HD 209458b is at level comparable to, or even greater than, the precision of the measurements for HD 189733b.  In other words, if that same precision could be achieved for HD 209458b, the magnetic effects would be observable, to the degree that they would need to be included in any models used to interpret the observations.

In addition to the information available from an orbital phase curve, the method of eclipse mapping can be used to reconstruct two-dimensional maps of the emission from the day side of a planet, by carefully measuring the shape of the light curve as the planet passes into and out of secondary eclipse.  Recently two groups have used this method to create maps of the 8\micron~emission from the day side of HD 189733b \citep{deWit2012,Majeau2012}, for the first time providing information about the latitudinal profile of emission.  Both groups confirmed a longitudinal offset of the hot spot that is consistent with the phase curve measurements, but their separate analysis approaches lead to a difference in their latitudinal profile results.  \citet{deWit2012} found that the hot spot was shifted north of the equator by $17\pm10$\degrees, while \citet{Majeau2012} combined the eclipse mapping and phase curve information to arrive at a value consistent with no latitudinal shift away from the equator ($3.1\pm9.4$\degrees~N).  A northward shift of the hottest region of the atmosphere would be surprising, given that most models of hot Jupiter circulation predict strong hemispheric symmetry for the temperature structure \citep{Showman2009,Heng2011,DobbsDixon2012,RM12b}.  Models that have broken north-south symmetry are also those that exhibit temporal variability \citep{Cho2003,Cho2008,Langton2007,DobbsDixon2010} and those able to produce the largest latitudinal shifts are also the most variable \citep{Rauscher2007} and would exceed the limit placed by \citet{Agol2010}.

As we have already discussed, magnetic effects are unlikely to have any significant influence on the atmosphere of HD 189733b, but it is worth considering whether they could in general produce observable latitudinal shifts of the hot spot.  Upon inspection of several snapshots from each of our models, we find strong hemispheric symmetry for all models of HD 189733b, as expected.  However, the $B=3$ and 10 G models of HD 209458b do show variation in the latitude of the hot spot, with offsets of up to $10-20$\degrees~N (or S).  However, those same models lack any significant offset of the hottest region from the substellar \emph{longitude}.  We find that magnetic effects cannot simultaneously account for significant offsets away from the substellar point in both latitude and longitude.

\subsection{Global ohmic heating rates} \label{sec:radii}

For many years now it has been recognized that the radii of hot Jupiters should exceed that of Jupiter because the intense stellar irradiation will slow their evolutionary cooling and contraction.  However, many transiting planets still have radii larger than expected, implying an additional source of heating that keeps some planets bloated.  Several mechanisms have been proposed to explain these inflated radii \citep[see][for a short review]{Fortney2010a}, but so far none are able to explain the full set of observed planets.  Ohmic heating is one of the candidates most recently proposed to provide the extra source of heating, by tapping into the kinetic energy of the atmospheric winds (generated by stellar heating) and transporting it deeper into the planet through the penetration and dissipation of induced currents \citep{Batygin2010,Perna2010b}.  The success of ohmic heating in producing inflated radii depends on how much power it can generate and at what depth, with deeper heating able to have a stronger effect on the planet's evolution \citep{Guillot2002}.

One of the difficulties in modeling the evolutionary impact of ohmic heating is that it couples regions of the atmosphere with short and long timescales (high in the atmosphere and the deep interior).  Another is that the one-dimensional models necessary for evolutionary calculations must use simple analytic forms to include the inherently three-dimensional nature of atmospheric circulation.  \citet{Batygin2011}, \citet{Wu2012}, and \citet{Huang2012} all calculate evolutionary models for hot Jupiters that include ohmic heating.  They each employ different assumptions, forms for the wind and temperature profiles, and prescriptions for the effects of feedback (such as magnetic drag slowing the winds, and heating of the upper atmosphere), and arrive at different conclusions regarding the effectiveness of ohmic heating as a mechanism for explaining the observed distribution of hot Jupiter radii.

\subsubsection{Our calculated heating rates}

\begin{figure}[ht!]
\begin{center}
\includegraphics[width=0.65\textwidth]{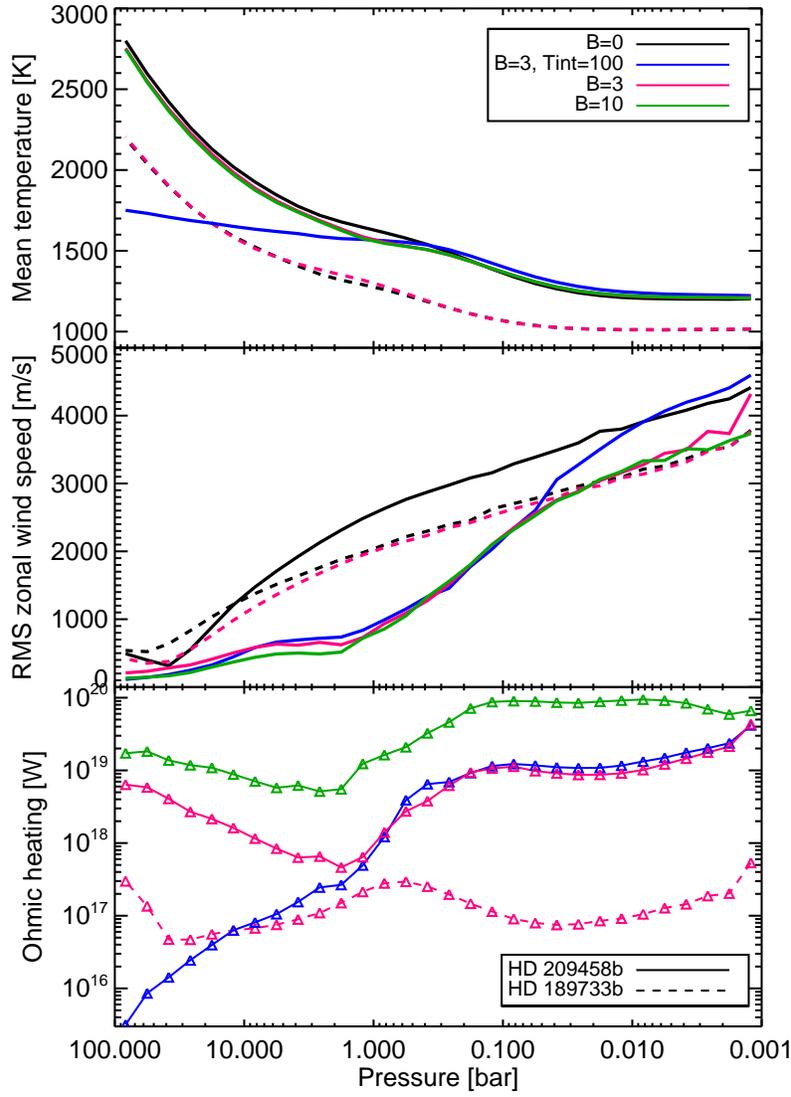}
\end{center}
\caption{Profiles of temperature, zonal wind speed, and ohmic heating as a function of pressure, for several of our models.  The temperature profile is the arithmetic mean at each pressure level and the wind profile is the root-mean-square.  The ohmic heating profile gives the total heating rate at each pressure level (see Appendix A for the integration), such that adding the contribution from each layer (each point) allows for visual integration above or below some level of interest.} \label{fig:pprofs}
\end{figure}

Using our models we can calculate the ohmic heating rates directly from the wind and temperature structure of the atmosphere, consistently including the influences of magnetic drag and heating on that atmospheric structure.  By continuously updating the local resistivities throughout the atmosphere, our model is able to predict the ohmic heating without any use of a prescribed form for the drag or wind profile.  In Table~\ref{tab:obs} we report the total ohmic heating in each model, summed over the entire domain of the simulations, from 1 mbar to 100 bar.  In Figure~\ref{fig:pprofs} we plot global profiles of temperature, wind speed, and ohmic heating as a function of pressure, for several of our models.  These profiles both demonstrate the interplay between various effects and can be used as a guide or boundary condition for one-dimensional evolutionary models.

As we saw in Section~\ref{sec:hd1}, the cooler temperatures throughout the atmosphere of HD 189733b result in weaker magnetic effects, and there is little difference between the non-magnetic and magnetic models of this planet.\footnote{We only show the $B=0$ and 3 G models in Figure~\ref{fig:pprofs}, but the temperature profiles for models up to $B=30$ G are nearly identical, and the wind speeds in the $B=30$ G model are decreased by only a few hundred meters per second.}  Although the HD~189733b $B=3$ G model has faster winds than the HD 209458b $B=3$ G model, its lower atmospheric temperatures result in ohmic heating rates about two orders of magnitude less than those for HD 209458b.  However, when we frame the total ohmic heating rates as efficiencies (relative to the stellar input, $\varepsilon$, reported in Table~\ref{tab:obs}), we find that some of the models of HD 189733b have heating efficiencies that are comparable to those of the HD~209458b models.  In particular, the models of HD 189733b with $B=30$ G have efficiencies similar to that of the $B=10$ G model of HD 209458b.  Nevertheless, evolutionary models that include ohmic heating find that, for the same heating efficiency, planets with lower effective temperatures have smaller radii and Jupiter-mass planets with $T_{\mathrm{eff}}\lesssim1400$ K experience no significant radius inflation at all \citep{Batygin2011}.  This is consistent with the observed radius of HD 189733b, which is not larger than predicted by evolutionary models with no extra source of heating.

The global profiles for our HD 209458b models show that most of the planet's ohmic heating occurs high in the atmosphere, where the winds are fast.  From previous sections we also know that most of this heating occurs on the hot day side (e.g., Figure~\ref{fig:r611}), where the magnetic timescales are the shortest.  This is an important point because it means that the globally averaged temperature profile is not representative of the temperatures at which the heating occurs.  If one-dimensional models were to use a globally averaged temperature profile to calculate ohmic heating, they would overestimate the atmospheric resistivities and underestimate the heating rates.

The $B=3$ and 10 G models have very similar temperature and wind profiles; the factor of $\sim$6 increase in ohmic heating for the $B=10$ G model (integrated over the entire atmosphere) is mainly due to the increased magnetic field strength.  Although magnetic effects do not significantly change the global temperature profiles, compared to the $B=0$ G model, they do strongly suppress wind speeds, by $\sim$1 \kms.  The greatest wind reduction occurs for pressures around 1 bar, although the winds are slowed as deep as $\sim$20 bar.  The strong decrease in winds leads to a dip in ohmic heating at $\sim$1 bar, while the heating below this level depends on the temperature structure of the deep atmosphere.  

\begin{figure}[ht!]
\begin{center}
\includegraphics[width=0.7\textwidth]{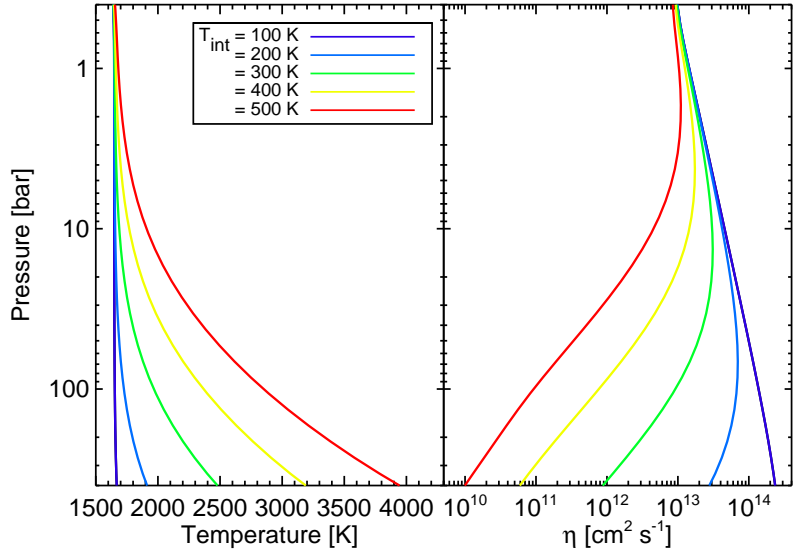}
\end{center}
\caption{Left: the analytic globally averaged temperature profiles for models of HD 209458b with a range of internal heat fluxes, characterized by values for \Tint~ranging from 100 to 500 K.  Right: the corresponding profiles of electrical resistivity ($\eta$), calculated as per Equation~\ref{eqn:eta}.} \label{fig:deepatm}
\end{figure}

The ohmic heating profiles for the $\Tint=100$ K and $\Tint=500$ K versions of our $B=3$~G HD~209458b model rapidly diverge for pressures greater than 1 bar.  The heating rate in the $\Tint=100$ K model drops by 2 orders of magnitude from 1 to 100 bar, while the heating rate in the $\Tint=500$ K model increases by an order of magnitude over the same pressure range.  The winds in the $\Tint=100$ K model are slower than in the $\Tint=500$ K model, 100 \ms~instead of 200 \ms, but the drastic reduction in ohmic heating is much more strongly dependent on the huge temperature difference.  In Figure~\ref{fig:deepatm} we plot analytic, globally averaged temperature-pressure profiles for HD 209458b models with $\Tint$ from 100 to 500 K, as well as the corresponding resistivity profiles.  The temperature profiles use the analytic formalism of \citet{Guillot2010}, appropriate for our models, and the resistivities are calculated from the temperature profiles using Equation~\ref{eqn:eta}.  Not only is there a difference of three orders of magnitude between the resistivities of the $\Tint=100$~K and 500 K models at 100 bar (the lower boundary of our simulations), but these profiles also show opposite behavior: from 10 to 100 bar (and deeper) the resistivities of the $\Tint=100$ K are increasing, while the resistivities of the $\Tint=500$ K model quickly decrease.  This difference leads to the disparate behavior of the ohmic heating profiles at depth.  The current value of \Tint~for HD~209458b (or any extrasolar planet) is unknown, meaning that the huge difference between these two models---a change in the ohmic heating at depth of at least two orders of magnitude---is a quantitative demonstration of our ignorance.

The internal heat flux of a planet should decrease as it ages and cools, so in some sense the comparison between these two models could be seen as between earlier ($\Tint=500$ K) and later ($\Tint=100$ K) stages of a planet's evolution, and we would therefore expect to find more ohmic heating at depth earlier in a planet's lifetime.  Note that our results do not necessarily contradict the assumption, used in \citet{Batygin2011} and \citet{Wu2012}, that the magnetic efficiency is constant throughout a planet's evolution.\footnote{While \citet{Huang2012} do not assume a constant efficiency, their models have ohmic heating at depth that is not monotonic with the inner entropy and it is difficult to make a direct comparison between their models and ours.}  Both studies constrain the total heating throughout the atmosphere to be constant, but allow for variation of the depth at which heating occurs.  In fact, due to the slightly higher temperatures and wind speeds in the upper atmosphere of our $\Tint=100$~K model, it has more ohmic heating at altitude than the $\Tint=500$ K model.  This compensates for the decrease in heating at deep pressures and results in these two models having the same total rate of ohmic heating (see Table~\ref{tab:obs}).  Since we only have this single example of identical models with different $\Tint$, we cannot comment on whether this is somehow a fundamental property, or just coincidence.

\subsubsection{Our findings on ohmic heating as a cause for HD 209458b's inflated radius}

Finally, we consider whether the amount of ohmic heating in any of our HD 209458b models is sufficient to explain this planet's bloated radius.  According to \citet{Guillot2002}, the inflated radius of HD 209458b requires roughly 10\% of the stellar insolation to be deposited as heating around the 5 bar level, $\varepsilon=1$\% if around 20 bar, or $\varepsilon \approx 0.08$\% if in the adiabatic interior, which begins at $\sim$160 bar in their model.  On the other hand, the results of \citet{Batygin2010} are able to reproduce the radius of HD 209458b with $4\times 10^{18}$ W of ohmic heating (or $\varepsilon \approx 0.01$\%) in the interior, which in their model begins at $\sim$90 bar, for the case of solar metallicity, no core, and $T_{\mathrm{iso}}=1700$ K.  Note that the amount of heating required depends not just on the pressure at which the heating is deposited, but also on the details of the particular model, including the metallicity of the atmosphere, the presence or absence of a core, the location of the radiative-convective boundary, and how all these vary throughout the planet's evolution \citep[see][]{Batygin2011,Huang2012,Wu2012}.  This motivates future work to develop a more consistent connection between ohmic heating results from evolutionary and circulation models.

Our $B=3$ G model cannot fulfill the \citet{Guillot2002} requirement of $\varepsilon=1$\% at 20 bar, nor can the $B=10$ G model, whose ohmic heating, integrated from 10 to 100 bar, only amounts to $\sim$0.2\% of the stellar insolation.  Although the total ohmic heating in this model is 4\% of the stellar heating, most of it is deposited at altitude and cannot effect the planet's evolution.  Due to the use of constant absorption coefficients in our radiative transfer scheme, the temperature profiles in our model will never become adiabatic (see Appendix B of RM12) and so we have no radiative-convective boundary.  If we consider the ohmic power in the deepest level of our models, which spans $\sim70-100$ bar, then both our $B=10$ G and $B=3$  ($\Tint=500$ K) models easily fulfill the \citet{Batygin2010} requirement of $4\times 10^{18}$ W at $\sim$90 bar.  However, the $B=3$~G model with $\Tint=100$ K is orders of magnitude away from reaching that heating threshold, again demonstrating the importance of the deep thermal structure on the amount of ohmic heating.

We can also employ the scalings from \citet{Wu2012} to estimate the ohmic power dissipated below our models' bottom boundary.  \citet{Wu2012} argue that the conservation of currents implies simple scalings between the radial and meridional components of atmospheric currents and those in the interior.  They make the case, based on conservation of total current, that at an order of magnitude level the total heating in the interior must be $\sim z_{\mathrm{wind}}/R_p$ smaller than in the ``weather layer'' (which has a vertical extent $z_{\mathrm{wind}}$).  The average thickness of our modeled atmosphere, from 100 bar to 1 mbar, is $\sim 2 \times 10^6$ m for all of our HD 209458b models.  From $z_{\mathrm{wind}}/R_p=0.02$ and the total ohmic heating rates in each of our models we find that the heating below our bottom boundary should be $\sim 6 \times 10^{18}$ W for $B=3$ G and $\sim 3 \times 10^{19}$ W for $B=10$~G.  Both estimates are above the $4\times 10^{18}$ W threshold from \citet{Batygin2010} and the $B=10$ G model matches the interior heating constraint from \citet{Guillot2002}.  (It is worth noting that our heating rates are not too dissimilar from the solar composition models of \citet{Batygin2010} with $T_{\mathrm{iso}}=1700$ K, with a few percent for the total ohmic heating  efficiency but only about a few times $10^{19}$ W dissipated deeper than 10 bar.)
From the scalings of \citet{Wu2012}, we would calculate identical interior dissipation rates for the $\Tint=100$ K and $\Tint=500$ K models with $B=3$ G, although we have shown that ohmic heating at depth depends very strongly on the deep atmosphere profile.  This is clearly a limitation of their idealized model, which assumes a constant conductivity from the upper to deep atmosphere, with the conductivity only changing at the transition to the interior.  Again, this is an issue that would benefit from more complete models that couple the atmosphere and interior.

The estimate from the scalings in \citet{Wu2012} for our interior heating rates rests on radial currents alone, while the formalism we use for the drag only describes meridional currents (which are dominant in the thin induction region).  In this way the theories are complementary, since meridional and radial currents are additive in terms of ohmic dissipation.  Therefore, our estimates from this scaling argument, together with the rising ohmic power at depth in our models (based on meridional currents), both suggest that interior ohmic dissipation can inflate HD 209458b for planetary magnetic field strengths of $B\geq 3-10$ G.

\section{Summary} \label{sec:conc}

We have updated our three-dimensional model of atmospheric circulation to include geometrically and energetically consistent magnetic drag and ohmic heating.  This is the first model to directly couple the magnetic effects to the full atmospheric structure.  We calculate the electrical resistivities directly from local conditions and update these values at each timestep in the simulation.  We apply magnetic drag only to the zonal (east-west) component of the flow and include the latitudinal dependence in the strength of the drag.  All kinetic energy lost through magnetic drag is consistently returned to the atmosphere as localized ohmic heating.  In these ways the magnetic effects are strongly coupled with the atmospheric circulation.  Here we present results from this code for two well known hot Jupiters, HD 189733b and HD 209458b.  We test models of these planets with planetary magnetic field strengths ranging from 0 to 30 G.

Due to the $\sim$300 K difference in equilibrium temperature between HD 189733b and HD~290458b, magnetic drag and ohmic heating only influence the circulation of the hotter planet, HD 209458b.  The $B=3$ G model of HD 209458b is obviously different from the non-magnetic version, but even the $B=30$ G model of HD 189733b with enhanced metallicity does not show significant differences from the $B=0$ G model, having only slightly slower wind speeds.  The models of HD 189733b also do not show any signature of magnetic effects in their observable properties, nor do they produce enough ohmic heating to effect the planet's evolution, which is consistent with the observed (non-inflated) radius of this planet.

We find that magnetic effects are able to strongly influence the circulation of HD 209458b; in several aspects the models with $B>0$ differ from the $B=0$ version.  In particular, the magnetic models: 1) have slower wind speeds, by $\sim$1 \kms (although still supersonic), 2) do not have an equatorial eastward jet that circles the globe, 3) have departures from hemispheric symmetry in the temperature and flow patterns, and 4) maintain more of a hot-day/cold-night temperature structure, over a deeper range of pressures, than the non-magnetic model.  These trends gradually become stronger at higher magnetic field strengths.  We also find that it is the magnetic drag that has a stronger influence on the circulation than the (coupled) ohmic heating, both of which act primarily on the day side of the planet.  Throughout almost all of the atmosphere the local ohmic heating is a very small fraction of the radiative heating.  However, our $B=30$ G model of HD 209458b has localized features where the ohmic heating reaches as much as 10\% of the local radiative heating; this model became numerically unstable and crashed before completion.

Magnetic drag and heating have a strong enough effect on the circulation of HD 209458b that they alter its observable properties.  The flux contrast between the day and night sides of the planet is greater in the magnetic models; during the planet's orbit the minimum flux emitted is only $12-13$\% of the maximum flux, compared to a ratio of 17\% for the non-magnetic model.  We also find that the brightest region of the atmosphere remains well aligned with the substellar longitude, compared to a 12\degrees~eastward shift in the $B=0$ G model.  Although there is no shift in longitude in the magnetic models, we do find that the latitude of the brightest region varies in time and can shift away from the substellar point by as much as $10-20$\degrees.  The differences between the magnetic and non-magnetic models of HD~209458b are at a level such that they could be measured, meaning that models used to interpret observations of this planet should include magnetic effects.

Finally, we compare the ohmic heating profiles from our models of HD 209458b to predictions from evolutionary models to determine whether we find sufficient heating at depth to explain the inflated radius of this planet.  Most of the ohmic heating in our models occurs high in the atmosphere and cannot prevent the planet's standard cooling and contraction.  However, the heating in the deepest layers of our $B=3$ and 10 G models could fulfill the requirement for inflation set by \citet{Batygin2010}, although only if the internal heat flux is high enough for a hot deep atmosphere; we found heating rates at depth to be two orders of magnitude higher in the $B=3$~G model with $\Tint=500$ K than the one with $\Tint=100$ K.  The models with $\Tint=500$ K also have heating profiles that are increasing with pressure from 10 to 100 bar, even though the winds are becoming slower, with speeds on the order of 100 \ms.  We use scaling arguments from \citet{Wu2012} to estimate the ohmic heating in the planet's interior, below our models' bottom boundary.  These estimates meet or exceed the requirements from \citet{Guillot2002} and \citet{Batygin2010} for ohmic heating in the adiabatic interior.  Both these estimates, and the rising ohmic power with pressure that we find in our models, suggest that interior ohmic dissipation can inflate the radius of HD 209458b for planetary magnetic field strengths at or greater than $B=3-10$ G.

In order to model the complex interaction between magnetic effects and the atmospheric circulation, we made several simplifying assumptions.  We expect these choices to be appropriate to first order, but as always, we would benefit from a fuller theory, in which these assumptions could be relaxed.  In particular, our calculation of the magnetic drag and ohmic heating is based on a formalism that assumes axisymmetry in the flow structure and atmospheric resistivities \citep{Liu2008}, neither of which is realized in hot Jupiter atmospheres, especially at low pressures.  We have also assumed the simplest magnetic geometry, with an aligned dipole field, while in reality the planet's magnetic axis could be misaligned from its axis of rotation, or the field could be multipole or uneven in other ways.  Varying the magnetic geometry could have strange impacts on the circulation.  In addition, our magnetic drag is only applied to the zonal flow, although the meridional flow may also begin to experience drag as it nears the poles and the field becomes more radial, but we lack the formalism to include this effect.  However, even with these simplifications, we have established that the inclusion of magnetic effects results in a greater richness of possibilities for atmospheric dynamics on the hottest exoplanets.  

\acknowledgements

We thank the anonymous referee for comments that helped to improve the clarity of this paper.  This work was performed in part under contract with the California Institute of Technology (Caltech) funded by NASA through the Sagan Fellowship Program.  ER thanks the Physics Department at Virginia Tech for kindly hosting her during the writing of this paper.  KM was supported by the NASA grant PATM NNX11AD65G.

\appendix

\section{An explanation of the different forms of ohmic heating shown in plots}

In Figures~\ref{fig:r611},~\ref{fig:r602}, and~\ref{fig:hd1b30} we plot the specific ohmic heating rate in units of W kg$^{-1}$.  These rates are calculated from the local atmospheric structure, following Equations~\ref{eqn:tmag} through~\ref{eqn:mheat}.

In Figure~\ref{fig:latprofs} we plot the ohmic heating rate as a function of latitude for various models.  Here we have calculated the heating rates everywhere in the model (as per above) and then summed over longitude and pressure at each latitude:
\begin{equation}
Q(\phi)/(d\phi) = \int (u^2/\tmag)\ dm/d\phi = \int_{P,\theta} (u^2/\tmag)\ (1/g) dP\ R_p^2 \cos (\phi)\ d\theta
\end{equation}
\noindent (where we have taken advantage of hydrostatic balance to convert $dm=\rho dz dA$ to pressure coordinates).  The resulting heating rates are in units of watts per radian.

In Figure~\ref{fig:pprofs} we plot the total ohmic heating at each pressure level, in units of watts:
\begin{equation}
Q (P) = (1/g) \Delta P \int_{\phi,\theta} (u^2/\tmag)\ R_p^2 \cos(\phi)\ d\phi \ d\theta \label{eqn:qp}
\end{equation}
\noindent where $\Delta P$ is the (model-dependent) thickness of each pressure level.

\end{document}